\documentclass[12pt,a4paper]{article}
\usepackage[utf8x]{inputenc}
\usepackage[T1]{fontenc}
\usepackage{amsfonts}
\usepackage{amsmath}
\usepackage[pdftex]{graphicx} 
\usepackage[pdftex,linkcolor=black,pdfborder={0 0 0}]{hyperref} 
\usepackage{calc} 
\usepackage{enumitem} 
\usepackage{amsthm}
\usepackage{bm}
\usepackage{xcolor}
\usepackage{natbib}
\usepackage{placeins}
\usepackage{comment}
\usepackage{algorithm2e}
\RestyleAlgo{ruled}
\usepackage{dsfont}
\usepackage{pdflscape}
\usepackage{pdfpages}

\def\spacingset#1{\renewcommand{\baselinestretch}%
{#1}\small\normalsize} \spacingset{1}

\DeclareMathOperator{\UD}{U}
\DeclareMathOperator{\ND}{\mathcal N}

\DeclareMathOperator{\GD}{\mathcal{G}}

\DeclareMathOperator{\IWD}{\mathcal{IW}}
\DeclareMathOperator{\WD}{\mathcal{W}}
\DeclareMathOperator{\HalfTD}{Half-t}
\newcommand\norm[1]{\lVert#1\rVert}
\newcommand\indicator[1]{\mathds{1}_{\left\{ #1 \right\}}}

\DeclareMathOperator{\vech}{\text{vech}}
\DeclareMathOperator{\vect}{\text{vec}}
\newcommand\KL[2]{\mathcal{D}_\text{KL}\left\lbrack #1 \,\vert\vert\, #2\right\rbrack}
\DeclareMathOperator{\diag}{diag}

\newtheorem{lemma}{Lemma}

\allowdisplaybreaks

\begin{document} 

\begin{titlepage}

\title{Variational inference for hierarchical models with conditional scale and skewness corrections}

\author{Lucas Kock$\mbox{}^{1\ast}$, Linda S. L. Tan$\mbox{}^1$, Prateek Bansal$\mbox{}^2$ and David J. Nott$\mbox{}^1$} 

\date{\today}
\maketitle
\thispagestyle{empty}
\noindent

\begin{center}
{\Large Abstract}
\end{center}
\vspace{-1pt}
\noindent Gaussian variational approximations are widely used for summarizing posterior distributions in Bayesian models, especially in high-dimensional settings. However, a drawback of such approximations is the inability to capture skewness or more complex features of the posterior. Recent work suggests applying skewness corrections to existing Gaussian or other symmetric approximations to address this limitation. We propose to incorporate the skewness correction into the definition of an approximating variational family. We consider approximating the posterior for hierarchical models, in which there are ``global'' and ``local'' parameters.  A baseline variational approximation is defined as the product of a Gaussian marginal posterior for global parameters and a Gaussian conditional posterior for local parameters given the global ones.  Skewness corrections are then considered. The adjustment of the conditional posterior term for local variables is adaptive to the global parameter value.    
Optimization of baseline variational parameters is performed jointly with the skewness correction. Our approach allows the location, scale and skewness to be captured separately, without using additional parameters for skewness adjustments.  The proposed method substantially improves accuracy for only a modest increase in computational cost compared to state-of-the-art Gaussian approximations. Good performance is demonstrated in generalized linear mixed models and multinomial logit discrete choice models. 

\vspace{20pt}
 
\noindent
{\bf Keywords}: hierarchical model, posterior skewness, skew-symmetric distribution, variational Bayes.

\vspace*{\fill}
\noindent {\small\textbf{Acknowledgments:} David Nott's research is supported by the Ministry of Education, Singapore, under the Academic Research Fund Tier 2 (MOE-T2EP20123-0009), and he is affiliated with the Institute of Operations Research and Analytics at the National University of Singapore. Linda Tan's research is supported by the Ministry of Education, Singapore, under its Academic Research Fund Tier 2 (MOE-T2EP20222-0002).} 

\vspace{20pt}

\noindent{\small
$^1$ Department of Statistics and Data Science, National University of Singapore\\
$^2$ Department of Civil and Environmental Engineering, National University of Singapore\\
$^\ast$ Correspondence should be directed to lucas.kock@nus.edu.sg
}

\end{titlepage}

\spacingset{1.5} 

\section{Introduction}

Gaussian variational approximations \citep{opper+a09} are popular for efficiently estimating the posterior distribution in Bayesian models with a high-dimensional parameter.  If the covariance matrix is sparsely parametrized, optimizing the approximation is computationally efficient even in large-scale problems.  This tractability is an important advantage. However, a disadvantage of Gaussian approximations is their inability to capture skewness and other complex features of the posterior. Several authors have developed variational inference approaches based on simple extensions of the Gaussian family to allow skewed approximations while maintaining computational feasibility.  Examples of this are approximations based on skew normal families \citep{ormerod11} and Gaussian copulas \citep{han2016}.

An alternative approach to capturing posterior skewness has been discussed recently by \cite{PozDurSza2024}, who develop an adjustment that can be applied post-hoc to any existing Gaussian or other symmetric approximation. The existing approximation  can be constructed using any convenient method,  such as variational approximation, Laplace approximation \citep{rue+mc09} or expectation propagation \citep{minka01}. The corrected approximating distribution has a closed form expression for its density, and it is possible to simulate from it directly using a rejection-free sampling scheme \citep{WanBoyGen2004}. This approach has both finite sample and asymptotic theoretical support under appropriate assumptions, and performs well in applications while being easy to implement.

Here we build on the method of \cite{PozDurSza2024} to improve standard Gaussian or conditionally structured Gaussian \citep{TanBhaNot2020} variational approximations.  We make three main contributions. First, we implement skewness corrections in hierarchical models where there are ``global'' as well as ``local'' parameters. We begin with a baseline Gaussian variational distribution that can be decomposed into a marginal posterior for global parameters and a conditional posterior for local parameters given the global ones. A skew-symmetric correction of the baseline variational approximation similar to \cite{PozDurSza2024} is then incorporated into the definition of the variational family. The baseline variational parameters are optimized jointly with the skewness correction, and the adjustment in the local conditional term is adaptive to the values of the global parameters. Optimization of the new variational family is more complex than in the Gaussian case, and our second contribution is to develop effective methods for stochastic gradient variance reduction in this context. The sampling scheme for simulating from the skewness-corrected variational density involves thresholding variables uniformly distributed on the unit cube, which introduces discontinuity. As such, the standard reparametrization trick \citep{KinWel2014,rezende+mw14} is inapplicable, and we propose a marginalization over the uniform variables, similar to the strategies in \cite{NaeRuiLinBle2017}, that enables the use of reparametrization gradients in the rejection-free sampling scheme. The third contribution is to demonstrate the effectiveness of our proposed approach in several applications using generalized linear mixed models and multinomial logit discrete choice models. Our methodology is shown to be more effective than simply applying the correction to a learnt approximation. However, computational efficiency is maintained as no additional variational parameters have to be introduced for implementing the skewness correction.  

Research in developing extensions of Gaussian variational families, which capture posterior skewness while retaining the tractability of Gaussian approximations is actively ongoing. Skew normal variational approximations based on the formulation of \cite{azzalini199} have been considered by \cite{ormerod11} who use one-dimensional quadrature to evaluate the variational lower bound and its gradients, \cite{lin+ks19} who develop natural gradient variational inference for mixture of exponential family approximations, and \cite{zhou+go23} who rely on matching key statistics of the posterior distribution. \cite{smith+ln20} and \cite{smith+l21} use Gaussian and skew normal copulas with simple marginal transformations to induce skewness, while \cite{dutta+vr25} obtain skew marginals using modified Laplace approximations combined with a Gaussian copula. To capture posterior conditional independence structure for hierarchical models, \cite{SalYuNotKoh2024} consider skew decomposable Gaussian graphical models \citep{zareifard+rkl16} as variational approximations. \cite{tan+c24} use a subclass of the closed skew normal densities \citep{gonzalez-farias+dg04}, which is constructed using affine transformations of independent univariate skew normals, as the variational family. Their approach falls under the affine independent variational framework of \cite{challis+b12}. \cite{fasano+dz22} consider a partially factorized approximation belonging to the class of unified skew normal distributions in high-dimensional probit regression. Theoretical support for skew normal posterior approximations is given by \cite{durante+ps23}, who derive a skewed Bernstein-von Mises theorem. While the above literature focuses on extending Gaussian families in variational inference to obtain skewed posterior approximations, the present work concentrates on incorporating the approach of \cite{PozDurSza2024} into variational inference for hierarchical models, for which there is exploitable posterior conditional independence structure.  

In the next section, we briefly review variational inference and a conditionally structured Gaussian variational approximation for hierarchical models discussed in \cite{TanBhaNot2020}.  We then describe the method of \cite{PozDurSza2024} for skewness correction of an existing Gaussian or other symmetric approximation. Section~\ref{sec:our} incorporates the method of \cite{PozDurSza2024} into the variational inference approach of \cite{TanBhaNot2020}, where the adjustment is made at the level of both a marginal global posterior term and a local conditional posterior term.  For the latter, the skewness adjustment is adaptive to the conditioning value for the global parameter. Section~\ref{sec:optimization} explains how to optimize the new variational family that incorporates skewness correction, and discusses variance reduction of stochastic gradients using a variation of the reparametrization trick.  Section~\ref{sec:finite_sample} discusses finite sample guarantees, and Section~\ref{sec:experiments} demonstrates the effectiveness of our method for improving posterior approximations in generalized linear mixed models and a random parameter multinomial logit model for a discrete choice experiment. Section~\ref{sec:discussion} concludes with some discussion. The code to reproduce the results in this article and to apply the proposed variational approximation to other models is publicly available at \href{https://github.com/kocklucx/GLOSS-VA}{github.com/kocklucx/GLOSS-VA}.

\section{Variational approximation for hierarchical models}\label{sec:review}

Throughout this article, we consider a hierarchical model with global parameters $\theta_G$ and latent variables $\theta_L=(b_1^\top\dots,b_n^\top)^\top$ for observations $y=(y_1^\top,\dots,y_n^\top)^\top$, where $y_i=(y_{i1},\dots,y_{in_i})^\top$. Here, $b_i$ is an observation-specific latent variable for $y_i$. Let $\theta=\left(\theta_L^\top,\theta_G^\top\right)^\top$ denote the variables, and $d$ and $d_i$ denote the dimensions of $\theta_G$ and $b_i$ respectively, for $i=1,\dots,n$. The likelihood is $$p(y\mid\theta)=\prod_{i=1}^np(y_i\mid b_i,\theta_G),$$ and we consider the prior 
$$p(\theta)=p(\theta_G)\prod_{i=1}^np(b_i\mid\theta_G).$$
Thus the $(d+\sum_{i=1}^nd_i)$ dimensional posterior is \begin{align}\label{eq:posterior}
    p(\theta\mid y)\propto 
    p(\theta_G)\prod_{i=1}^n h_i(b_i\mid\theta_G),
\end{align}
where $h_i(b_i\mid\theta_G) = p(b_i\mid\theta_G)p(y_i\mid b_i,\theta_G)$ for $i=1, \dots, n$.

\subsection{Variational inference}\label{subsec:vi}
The goal of variational inference \citep[VI;~][]{BleKucMca2017} is to learn an approximation $q_\lambda(\theta)$ to the posterior $p(\theta\mid y) \propto p(y\mid\theta) p(\theta)=:h(\theta)$ by optimizing the variational parameters
$\lambda$. For instance, the approximation may be a multivariate normal, with $\lambda$ parametrizing the mean and covariance matrix.  The optimal value $\lambda^*$ of $\lambda$ is usually obtained by minimizing the reverse Kullback-Leibler (KL) divergence, 
\begin{equation}\label{eq:kl}
\KL{q_\lambda}{p(\cdot\mid y)}=\mathbb{E}_{q_\lambda(\theta)}\left[\log q_\lambda(\theta)-\log p(\theta\mid y)\right],
\end{equation}
where $\mathbb{E}_{q_\lambda(\theta)}[\cdot]$ denotes expectation with
respect to $q_\lambda(\theta)$.  It is straightforward to show  \citep[e.g.][]{OrmWan2010} that minimizing \eqref{eq:kl} is equivalent to maximizing the evidence lower bound (ELBO) given as $\mathcal{L}(\lambda)=\mathbb{E}_{q_\lambda(\theta)}\left[\log h(\theta)-\log q_\lambda(\theta)\right]$.
This objective can be conveniently optimized using stochastic gradient ascent (SGA) when the expectation cannot be evaluated in closed form. In SGA, the vector $\lambda$ is iteratively updated as $\lambda^{(t+1)}=\lambda^{(t)}+\rho^{(t)}\circ\widehat{\nabla_\lambda\mathcal{L}\left(\lambda^{(t)}\right)}$ at each iteration $t$. Here $\rho^{(t)}$ is a vector-valued step size, $\circ$ denotes element-wise multiplication, and $\widehat{\nabla_\lambda\mathcal{L}\left(\lambda^{(t)}\right)}$ is an unbiased estimate of the gradient of $\mathcal{L}(\lambda)$ evaluated at $\lambda^{(t)}$, using a sample drawn randomly from $q_{\lambda^{(t)}}(\theta)$. 

If a sample from $q_\lambda(\theta)$ can be generated by first drawing $\varepsilon$ from a distribution $\pi(\varepsilon)$, that does not depend on $\lambda$, and then transforming $\varepsilon$ by a deterministic function $\theta=f(\varepsilon,\lambda)$, then the ELBO can be written as an expectation with respect to $\pi(\varepsilon)$,
\begin{equation*}
    \mathcal{L}(\lambda)=\mathbb{E}_{\pi(\varepsilon)}\left[\log h(f(\varepsilon,\lambda))-\log q_\lambda(f(\varepsilon,\lambda))\right].
\end{equation*}
Thus,
\begin{equation}\label{eq:repgrad}
    \nabla_\lambda\mathcal{L}(\lambda)=\mathbb{E}_{\pi(\varepsilon)}\left[\nabla_\lambda\left\{\log h(f(\varepsilon,\lambda))-\log q_\lambda(f(\varepsilon,\lambda))\right\}\right].
\end{equation}
Estimating \eqref{eq:repgrad} with a single draw from $\pi(\varepsilon)$ yields an unbiased gradient estimate $\widehat{\nabla_\lambda\mathcal{L}\left(\lambda^{(t)}\right)}$, which usually has low variance, and this procedure is known as the reparameterization trick \citep{KinWel2014,rezende+mw14}.

\subsection{Conditionally structured Gaussian approximations}\label{subsec:csgva}
From the expression \eqref{eq:posterior} for the posterior, the local latent variables in $b$ are conditionally independent of each other given $\theta_G$, which motivates a variational approximation of the form,
\begin{align}\label{eq:q_structure}
q_\lambda(\theta)=q_\lambda(\theta_G)\prod_{i=1}^nq_\lambda(b_i\mid\theta_G).
\end{align}
If $q_\lambda(\theta)$ is chosen to be jointly Gaussian, this conditional independence structure can be captured by a sparse precision matrix \citep[G-VA;~][]{TanNot2018}. That is, $q_\lambda(\theta)=\varphi(\theta;\mu,\Sigma)$ with $\mu=\left(\mu_G^\top,m_1^\top,\dots,m_n^\top\right)^\top$ and $\Sigma^{-1} = TT^{\top}$, where
\begin{align*}
    T = \begin{pmatrix}
        T_1 & 0 & \cdots & 0&0\\
        0 & T_2 & \cdots & 0 &0\\
        \vdots& \vdots &\ddots&\vdots&\vdots\\
        0&0&\cdots&T_n&0\\
        T_{G1}&T_{G2}&\cdots&T_{Gn}&T_G
    \end{pmatrix}.
\end{align*}
Here, $\varphi(\cdot;\mu,\Sigma)$ denotes the density of the multivariate Gaussian with mean $\mu$ and covariance matrix $\Sigma$, $T_i$ and $T_G$ are lower triangular matrices with positive diagonals, of order $d_i$ and $d$ respectively, while $T_{Gi}$ is a $d\times d_i$ matrix, for $i=1,\dots,n$. From standard results on the conditional distributions of a multivariate Gaussian, we can derive 
\[
q_\lambda(\theta_G)=\varphi(\theta_G;\mu_G,T_G^{-\top}T_G^{-1}) 
\quad \text{and} \quad 
q_\lambda(b_i\mid\theta_G)=\varphi(b_i;\mu_i(\theta_G),T_i^{-\top}T_i^{-1}),
\] 
where $\mu_i(\theta_G)=m_i+T_i^{-\top}T_{Gi}^\top(\mu_G-\theta_G)$ is a function of $\theta_G$. The variational parameters to be optimized is thus 
$$\lambda=\left(\mu_G^\top,\vech(T_G),m_1^\top,\dots,m_n^\top,\vect(T_{G1}),\dots,\vect(T_{Gn}),\vech(T_1)^\top,\dots,\vech(T_n)^\top\right)^\top.$$
For a square matrix $A$, $\vect(A)$ is the vectorization of $A$ which stacks the columns into a vector from left to right, while $\vech(A)$ is the half-vectorization of $A$ which is obtained from $\vect(A)$ by omitting elements above the diagonal.

In the above variational approximation, the location of $b_i$ depends conditionally on $\theta_G$, but the covariance structure of $q_\lambda(b_i\mid\theta_G)$ is independent of $\theta_G$. As such a structure may be too restrictive in many applications,
\citet{TanBhaNot2020} propose a conditional scale correction to $q_\lambda(b_i\mid\theta_G)$, in which they express $T_i=T_i(\theta_G)$ as a function of $\theta_G$ as well. 
They call their approach conditionally structured Gaussian variational approximation (CSG-VA).  
Although $T_i(\theta_G)$ can be an arbitrary parametric function of $\theta_G$ in theory, \citet{TanBhaNot2020} consider 
\begin{equation}\label{eq:CSG_scale}
    \vech(T_i(\theta_G)^*)=f_i+B_i\theta_G,
\end{equation}
where $B_i$ is a ${d_i(d_i+1)}/2\times d$ matrix and $f_i$ is a vector of length ${d_i(d_i+1)}/2$. Here, we introduce the operator $\mbox{}^*$ for any matrix $A$, such that $A^*_{ij}=A_{ij}$ if $i\not=j$ and $A^*_{jj}=\log(A_{jj})$. Note that $f_1,\dots,f_n$ and $\vect(B_1),\dots,\vect(B_n)$ are additional variational parameters to be optimized. While this CSG-VA still assumes multivariate Gaussians for $q_\lambda(\theta_G)$ and $q_\lambda(b_i\mid\theta_G)$, $i=1,\dots,n$, the joint variational approximation $q_\lambda(\theta)$ is no longer Gaussian.

\subsection{Post-hoc skewness corrections}\label{subsec:posthoc_skewness}
Next, we review the skewness correction proposed by \citet{PozDurSza2024}. Let $q(\theta)=q^{(\mu)}(\theta)$ be a symmetric approximation of the posterior with symmetry point $\mu$, such as the Gaussian variational approximation introduced in \eqref{eq:q_structure}. Then $q^{(\mu)}(\theta) = q^{(\mu)}(2\mu - \theta)$ for any $\theta$ in the support of $q^{(\mu)}(\cdot)$. As $q^{(\mu)}(\theta)$ cannot capture any potential skewness in the true posterior $p(\theta\mid y)$, \citet{PozDurSza2024} propose a way to perturb the approximation post-hoc. They consider the class of skew–symmetric perturbations \citep{AzzCap2003,WanBoyGen2004},
\begin{align}\label{eq:skew_sym_distr}
    q^w(\theta)=2q^{(\mu)}(\theta)w(\theta),
\end{align}
where $w(\theta) \in [0,1]$ is a skewing function that satisfies $w(\theta)=1-w(2\mu-\theta)$ for the symmetry point $\mu$ of $q^{(\mu)}(\theta)$. To generate a sample from $q^w(\theta)$, we can use a rejection-free scheme by first drawing $\theta_{\text{temp}}$ from $q^{(\mu)}(\theta)$ and $u$ from $U[0,1]$, and then setting $\theta = \theta_{\text{temp}}$ if 
$u \leq w(\theta_{\text{temp}})$ and 
$\theta = 2\mu - \theta_{\text{temp}}$ if 
$u > w(\theta_{\text{temp}})$. 
Thus, a direct interpretation of \eqref{eq:skew_sym_distr} is that samples from $q^{(\mu)}(\theta)$ are reflected at $\mu$ with probability $1-w(\theta)$, so that $q^{w}(\theta)$ is skewed. 

For any posterior $\pi(\theta) = p(\theta \mid y)$, \citet{PozDurSza2024} show that a skew-symmetric representation of the form in \eqref{eq:skew_sym_distr} can be obtained by considering its symmetrized form about a point $\hat{\theta}$ in its support, $\bar{\pi}(\theta) = [\pi(\theta) + \pi(2\hat{\theta} -\theta)]/2$. Then $\pi(\theta) = 2 \bar{\pi}(\theta) w_{\hat{\theta}}  (\theta)$, where the skewing function is given by
\begin{align*}
w_{\hat{\theta}} (\theta) = \frac{\pi(\theta)}{2\bar{\pi}(\theta)} 
= \frac{h(\theta)/p(y)}{[h(\theta) + h(2\hat{\theta} - \theta)]/p(y)} 
= \frac{h(\theta)}{h(\theta) + h(2\hat{\theta} - \theta)},
\end{align*}
and $h(\theta)=p(\theta)p(y\mid\theta)$. They further prove that given a symmetric approximation $q^{(\mu)}(\theta)$, the optimal skewness correction minimizing $\KL{q^w(\theta)}{p(\theta\mid y)}$ is given by 
\begin{align}\label{eq:w_pozza}
    w^*(\theta)=\frac{h(\theta)}{h(\theta)+h(2\mu-\theta)}.
\end{align}
If the true posterior is symmetric about $\mu$, then $h(\theta)=h(2\mu-\theta)$ and $q^{w^*}(\theta)=q^{(\mu)}(\theta)$. There are no adjustable parameters to be determined in $w^*(\theta)$, as it only depends on $h(\theta)$ and the symmetry point $\mu$ of $q^{(\mu)}(\theta)$.

\section{Conditional scale and skewness corrections}\label{sec:our}
The novel variational family that we propose combines the CSG-VA of \citet{TanBhaNot2020} introduced in Section~\ref{subsec:csgva} with skewness corrections of the form in \eqref{eq:skew_sym_distr}. In contrast to \citet{PozDurSza2024}, we do not consider a global skewness correction. Instead, we apply the skewness corrections hierarchically by matching the structure of the true posterior. In addition, we optimize the variational parameters while taking the skewness correction into account, rather than only applying the correction post-hoc.  We call our novel method Global LOcal Scale and Skewness Variational Approximation (GLOSS-VA).  

Our variational family is of the form in \eqref{eq:q_structure} with
\begin{align*}
q_\lambda(\theta_G)&=2 \, \varphi(\theta_G;\mu_G,T_G^{-\top}T_G^{-1}) \, w(\theta_G),\\
q_\lambda(b_i\mid\theta_G)&=2 \, \varphi(b_i;\mu_i(\theta_G),T^{-\top}_i(\theta_G)T_i^{-1}(\theta_G)) \, w(b_i), \quad \text{for} \quad i=1,\dots,n,
\end{align*}
where $\mu_i(\theta_G)=m_i+T_i(\theta_G)^{-\top}T_{Gi}^\top(\mu_G-\theta_G)$ and $\vech(T_i(\theta_G)^*)=f_i+B_i\theta_G$ are as introduced in Section~\ref{subsec:csgva}. The skewing functions $w(\theta_G)$ and $w(b_i)$ for $i=1,\dots,n$  are described below. 

As $q_\lambda(b_i\mid\theta_G)$ is the variational approximation for $p(b_i\mid\theta_G)\propto h_i(b_i\mid\theta_G)$, following \citet{PozDurSza2024}, the optimal skewing function $w(b_i)$ is given by 
\begin{equation} \label{eq:w_l}
    w(b_i)=\frac{h_i(b_i\mid\theta_G)}{h_i(b_i\mid\theta_G)+h_i(2\mu_i(\theta_G)-b_i\mid\theta_G)},
    \quad 
    i=1, \dots, n,
\end{equation}
where $h_i(b_i\mid\theta_G) = p(b_i\mid\theta_G)p(y_i\mid b_i,\theta_G)$. Note that the symmetry point $\mu_i(\theta_G)$ is dependent on $\theta_G$, which is a situation that has not been considered in \citet{PozDurSza2024}.

Similarly, the optimal skewness correction  minimizing the Kullback-Leibler divergence between $q_\lambda(\theta_G)$ and $p(\theta_G\mid y)$ is given by
\begin{equation}\label{eq:w_G_bar}
    w(\theta_G)=\frac{{h}(\theta_G)}{{h}(\theta_G)+{h}(2\mu_G-\theta_G)},
\end{equation}
where ${h}(\theta_G)=p(\theta_G)p(y\mid\theta_G)$ is the kernel of the marginal posterior $\int p(\theta\mid y)d\theta_L$. Since $p(y \mid\theta_G)$ is in general not analytically available, we approximate
\begin{align*}
    {h}(\theta_G) &= p(\theta_G)\frac{p(\theta_L\mid\theta_G)p(y\mid\theta_L,\theta_G)}{p(\theta_L\mid\theta_G,y)}  \\
    &\approx p(\theta_G)\frac{ \prod_{i=1}^n h_i(b_i \mid \theta_G)}{\prod_{i=1}^n q_\lambda(b_i\mid\theta_G)}\\
    &\approx p(\theta_G)\prod_{i=1}^n\frac{h_i(\mu_i(\theta_G)\mid\theta_G)}{2 \varphi(\mu_i(\theta_G);\mu_i(\theta_G),T^{-\top}_i(\theta_G)T_i^{-1}(\theta_G)) \, w(\mu_i(\theta_G))}\\
    &=p(\theta_G)\prod_{i=1}^n \left[(2\pi)^{d_i/2}\det(T^{-\top}_i(\theta_G)T_i^{-1}(\theta_G))^{1/2}h_i(\mu_i(\theta_G)\mid\theta_G)\right] = \tilde{h}(\theta_G),
\end{align*}
where we set $w(\mu_i(\theta_G))=0.5$. Approximating ${h}(\theta_G)$ by $\tilde{h}(\theta_G)$ in the right-hand side of \eqref{eq:w_G_bar} enables us to compute $w(\theta_G)$. If $q(\theta_L\mid\theta_G,y)$ is a good approximation to $p(\theta_L\mid\theta_G,y)$, this approximation is tight. It's quality is further discussed in Supporting Information~B.

The full variational posterior including the local and global skewness correction is thus 
\begin{equation}\label{eq:q} 
q_\lambda(\theta) = 2 \, \varphi(\theta_G;\mu_G,\Sigma_G) \, w(\theta_G) \prod_{i=1}^n\left[2 \, \varphi(b_i;\mu_i(\theta_G),\Sigma_i(\theta_G)) \, w(b_i)\right].
\end{equation}
Due to the dependence of $T_i(\theta_G)$ on $\theta_G$ and the introduction of skewness through $w(\theta_G)$ and $w(b_i)$, $q_\lambda(\theta)$ can be highly non-Gaussian. Efficient sampling from \eqref{eq:q} can be accomplished as described in Algorithm~\ref{alg:sampling}. The full set of variational parameters to be optimized is
\begin{equation*}
\lambda=\left(\mu_G^\top,\vech(T_G^*),m_1^\top,\dots,m_n^\top,\vect(T_{G1}),\dots,\vect(T_{Gn}),f_1^\top,\dots,f_n^\top,\vect(B_1),\dots,\vect(B_n)\right)^\top.
\end{equation*}

\RestyleAlgo{ruled}
\begin{algorithm}[bt!]
\caption{Drawing a sample from $q_{\lambda}(\theta)$}\label{alg:sampling}
\KwIn{$\lambda$ (parameters of the variational approximation)}
Sample $\varepsilon_G\sim\ND(0,I_d)$\;
$\theta_G \leftarrow \mu_G+T_G^{-\top}\varepsilon_G$\;
Sample $u_G\sim\UD(0,1)$\;
\If{$u_G>w(\theta_G$)}{
    Set $\theta_G \leftarrow 2\mu_G-\theta_G$\;
}
\For{$i\gets1$ \KwTo $n$}{
Calculate $\mu_i\leftarrow\mu_i(\theta_G)$ and $T_i\leftarrow T_i(\theta_G)$\;
Sample $\varepsilon_i\sim\ND(0,I_{d_i})$\;
$b_i \leftarrow \mu_i+T_i^{-\top}\varepsilon_i$\;
Sample $u_i\sim\UD(0,1)$\;
\If{$u_i>w(b_i$)}{
    Set $b_i \leftarrow 2\mu_i-b_i$\;
}
}
Return $\theta=\left(b_1^\top,\dots,b_n^\top,\theta_G^\top\right)^\top$\;
\end{algorithm}

\section{Optimizing the ELBO}\label{sec:optimization}
Using the sampling procedure outlined in Algorithm~\ref{alg:sampling}, the ELBO $\cal L(\lambda)$ can be written as an expectation over $\pi(\varepsilon,u)=\pi(\varepsilon)\pi(u)$, where $\varepsilon=(\varepsilon_G^\top,\varepsilon_1^\top,\dots,\varepsilon_n^\top)^\top$ follows a standard Gaussian distribution and $u=(u_G,u_1,\dots,u_n)^\top$ is uniform on the unit hypercube. However, this is insufficient to employ the reparameterization trick, due to the discontinuity introduced by $u$ in the transformation $\theta=f(u,\varepsilon,\lambda)$ as implied in Algorithm~\ref{alg:sampling}. Instead, we consider
$$\mathcal{L}(\lambda)=\mathbb{E}_{\pi(\varepsilon,u)}[l(f(u,\varepsilon,\lambda))]=\mathbb{E}_{\pi(\varepsilon)}\left[\int l(f(u,\varepsilon,\lambda)) d\pi(u)\right],$$ 
where $l(\theta)=\log h(\theta)-\log q_\lambda(\theta)$, and the integral with respect to $\pi(u)$ can be solved analytically. To see this, consider a draw for $\theta_G$. Let $\tilde{f}_G(\varepsilon_G,\lambda)=\mu_G+T_G^{-\top}\varepsilon_G$. Then 
$$\theta_G=f_G(u_G,\varepsilon_G,\lambda)=\indicator{u_G\leq w(\tilde{f}_G(\varepsilon_G,\lambda))}\tilde{f}_G(\varepsilon_G,\lambda)+\indicator{u_G>w(\tilde{f}_G(\varepsilon_G,\lambda))}\left[2\mu_G-\tilde{f}_G(\varepsilon_G,\lambda)\right].$$ 
Therefore 
\begin{align*}
\int \tilde{l}(f_G(u_G,\varepsilon,\lambda))d\pi(u_G) 
& = w\left(\tilde{f}_G(\varepsilon_G,\lambda)\right)\tilde{l}\left(\tilde{f}_G(\varepsilon_G,\lambda)\right) \\
& \quad +\left(1-w\left(\tilde{f}_G(\varepsilon_G,\lambda)\right)\right)\tilde{l}\left(2\mu_G-\tilde{f}_G(\varepsilon_G,\lambda)\right),    
\end{align*}
for some function $\tilde{l}(\cdot)$.
A similar decomposition can be derived for the draws $b_i\mid\theta_G$ for $i=1,\dots,n$. Thus, $\int l(f(u,\varepsilon,\lambda)) d\pi(u)$ is a complex and nested combination of evaluations of $l(\theta)$ at different reflections for $\theta$, with weights depending on $\lambda$ and $\varepsilon$. A closed form derivation of $\int l(f(u,\varepsilon,\lambda)) d\pi(u)$ is given in Appendix~A.
We can thus write
\begin{align}\label{eq:our_repgrad}
    \nabla_\lambda\mathcal{L}(\lambda)=\mathbb{E}_{\pi(\varepsilon)}\left[\nabla_\lambda\int l(f(u,\varepsilon,\lambda)) d\pi(u)\right],
\end{align}
which has the same structure as \eqref{eq:repgrad}, and hence the reparameterization trick can be applied to derive an unbiased gradient estimate $\widehat{\nabla_\lambda\mathcal{L}(\lambda)}$. 
The idea to marginalize over $u$ is similar to the strategies proposed in \citet{NaeRuiLinBle2017}, where the task of combining the reparameterization trick with acceptance-rejection sampling is considered.

The complexity of the optimization depends on the structure of $h$ and therefore on the underlying statistical model. Initializing the stochastic gradient ascent at a good starting location as well as the randomness introduced by the reparameterization trick can be helpful in avoiding local optima. Starting the algorithm from different locations can also be helpful when the algorithm gets stuck in local optima. We however have not done so in our experiments as we have not experienced this issue in the examples considered. However, we acknowledge that this is a risk in all non-convex optimization problems.

\section{Finite sample guarantees}\label{sec:finite_sample}
Our novel GLOSS-VA has two main contributions, which differ from the method of \citet{PozDurSza2024}. First, we propose a hierarchical skewness correction that matches the conditional independence structure of the true posterior, thus allowing the application of separate skewness corrections to different blocks of parameters. Second, we do not apply the skewness correction post-hoc, but optimize the variational parameters while taking the skewness correction into account. To understand the individual impact of these contributions we consider the following alternatives to GLOSS-VA:

\begin{enumerate}
    \item[] G-VA: A Gaussian variational approximation with a  sparse precision matrix \citep[G-VA,~][]{TanNot2018};
    \item[] G-VA$^{G-}$: G-VA with the global skewness correction in \eqref{eq:w_pozza} by \citet{PozDurSza2024} applied post-hoc;
    \item[] G-VA$^{G+}$: G-VA with the global skewness correction \eqref{eq:w_pozza} learned with the variational parameters;
    \item[] G-VA$^{H-}$: G-VA with our novel hierarchical skewness correction applied post-hoc;
    \item[] CSG-VA: The conditionally structured Gaussian variational approximation by \cite{TanBhaNot2020};
    \item[] CSG-VA$^{H-}$: CSG-VA with our novel hierarchical skewness correction applied post-hoc.
\end{enumerate}

We now give a brief discussion of finite sample guarantees of improved performance for GLOSS-VA. Numerical experiments on real data are presented in Section~\ref{sec:experiments}. For each of the methods, $\bullet\in\{$G-VA, G-VA$^{G-}$, G-VA$^{G+}$, G-VA$^{H-}$, CSG-VA, CSG-VA$^{H-}$, GLOSS-VA$\}$, we write $\Lambda(\bullet)$ for the domain of the variational parameters and  denote the optimal parameter as $\lambda^\ast(\bullet)\in\Lambda(\bullet)$. Since neither the global nor the hierarchical skewness correction introduces additional parameters, we have 
\begin{equation*}
    \Lambda(\text{G-VA})=\Lambda(\text{G-VA$^{G-}$})=\Lambda(\text{G-VA$^{G+}$})=\Lambda(\text{ G-VA$^{H-}$}),
\end{equation*}
and similarly,
\begin{equation*}
    \Lambda(\text{CSG-VA})=\Lambda(\text{CSG-VA$^{H-}$})=\Lambda(\text{GLOSS-VA}).
\end{equation*}
As CSG-VA recovers the Gaussian approximation G-VA by setting $B_i=0$ in \eqref{eq:CSG_scale} for $i=1,\dots,n$, $\Lambda(\text{G-VA})\subseteq\Lambda(\text{CSG-VA})$. Applying a skewness correction post-hoc does not change the variational parameters, so that
$\lambda^\ast(\text{G-VA})=\lambda^\ast(\text{G-VA$^{G-}$})=\lambda^\ast(\text{ G-VA$^{H-}$})$, and $\lambda^\ast(\text{CSG-VA})=\lambda^\ast(\text{CSG-VA$^{H-}$})$. However, incorporating the skewness correction into the optimization may improve the parameters in the sense of Lemma~\ref{lem:1}. 

\begin{lemma}\label{lem:1}
Let ${\cal L}(\lambda)$ denote the ELBO under the GLOSS-VA variational family, then ${\cal L}(\lambda^\ast(\text{CSG-VA$^{H-}$}))\leq {\cal L}(\lambda^*(\text{GLOSS-VA}))$.
\end{lemma}

With proper optimization, GLOSS-VA can be no worse than CSG-VA with post-hoc correction, and G-VA$^{G+}$ improves on G-VA$^{G-}$ in the same sense. However, the optimization problem for GLOSS-VA is much more challenging than that for CSG-VA. Hence, it may be beneficial in practice to fit a simpler variational family and apply the skewness correction post-hoc. A more detailed study of theoretical guarantees of the GLOSS-VA method is an interesting topic for future work. 
This analysis also implies a clear ordering of the different variational families both in terms of their approximation and in computational complexity. This ordering is matched in numerical experiments as discussed in the next section. 

\section{Experiments}\label{sec:experiments}

We compare the performance of our novel GLOSS-VA method with the benchmarks described in Section~\ref{sec:finite_sample}. All approximations are evaluated using MCMC sampling as the gold standard. 

\subsection{Logistic mixed model}
First, we consider a longitudinal study on the health effects of air pollution \citep{FitLai1993}, which reports the wheezing status $y_{ij}$ of $n=537$ children annually from the ages 7 to 10 (\texttt{age}), for $i=1,\dots,n$, $j=1,\dots,4$. An additional covariate is the mother’s smoking status (\texttt{smoke}). We consider a random intercept logistic regression model,
\begin{equation*}
\log\left(\frac{\text{Pr}(y_{ij}=1)}{1-\text{Pr}(y_{ij}=1)}\right)=\beta_{0}+\texttt{smoke}_{ij}\beta_{\texttt{smoke}}+\texttt{age}_{ij}\beta_{\texttt{age}}+\left(\texttt{smoke}_{ij}\times\texttt{age}_{ij}\right)\beta_{\texttt{smoke}\times\texttt{age}}+b_i,
\end{equation*}
where $\beta=\left(\beta_{0},\beta_{\texttt{smoke}},\beta_{\texttt{age}},\beta_{\texttt{smoke}\times\texttt{age}}\right)^\top$ is the vector of fixed effects and $b_i\sim\ND(0,\exp(-2\eta))$ is an individual-specific random intercept, where $\eta$ is an unrestricted parameter corresponding to the log precision of the random effects prior. The global variables are $\theta_G=\left(\beta^\top,\eta\right)^\top$ and we consider the prior $\theta_G\sim\ND(0,10^2I_5)$.
This example is a popular benchmark in the VI literature \citep[e.g.][]{TanBhaNot2020,SalYuNotKoh2024} due to its complex posterior asymmetries, and we consider an alternative prior-structure in Supporting Information~D.1. 

\begin{figure}[tb]
    \centering
    \includegraphics[width=0.9\linewidth,keepaspectratio]{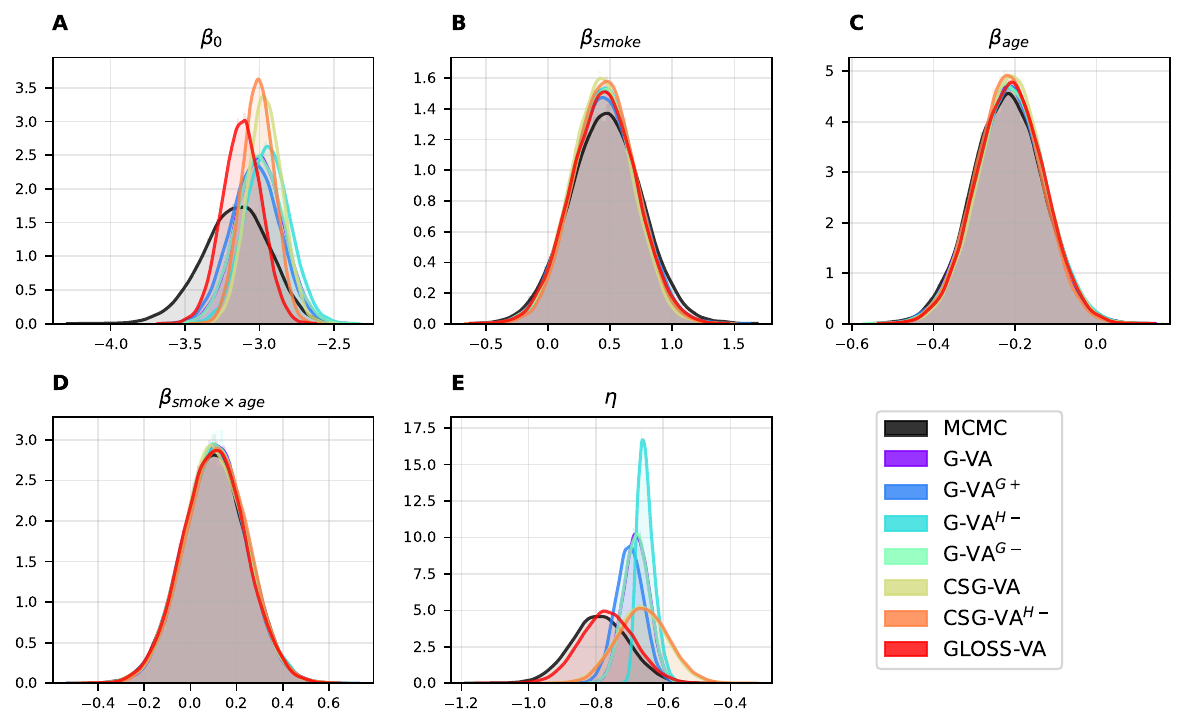}
    \caption{\small  Logistic mixed model. Univariate marginal posteriors for all global parameters. 
    }
    \label{fig:sixcities_global}
\end{figure}

Figure ~\ref{fig:sixcities_global} shows the approximated marginal posteriors for $\theta_G$. While all methods successfully approximate $\beta_{\texttt{smoke}},\beta_{\texttt{age}}$ and $\beta_{\texttt{smoke}\times\texttt{age}}$, GLOSS-VA shows improved performance in estimating the intercept $\beta_0$, and is the only method that captures the shape of $p(\eta\mid y)$ accurately.

Scatter plots displaying the samples drawn from $p(\beta_0,\eta\mid y)$ using MCMC and each approximation method are given in Appendix~C. The fixed intercept $\beta_0$ and $\eta$, which controls the variance of the random intercept, exhibit a complex dependence structure. This intricate relationship is approximated more accurately by CSG-VA, CSG-VA$^{H-}$ and GLOSS-VA, which suggests that the conditional scale correction is very helpful. Of all the approximation methods, GLOSS-VA matches the MCMC samples most closely and the resulting approximation is also clearly non-Gaussian.

\begin{figure}[tb]
    \centering
    \includegraphics[width=0.95\linewidth,keepaspectratio]{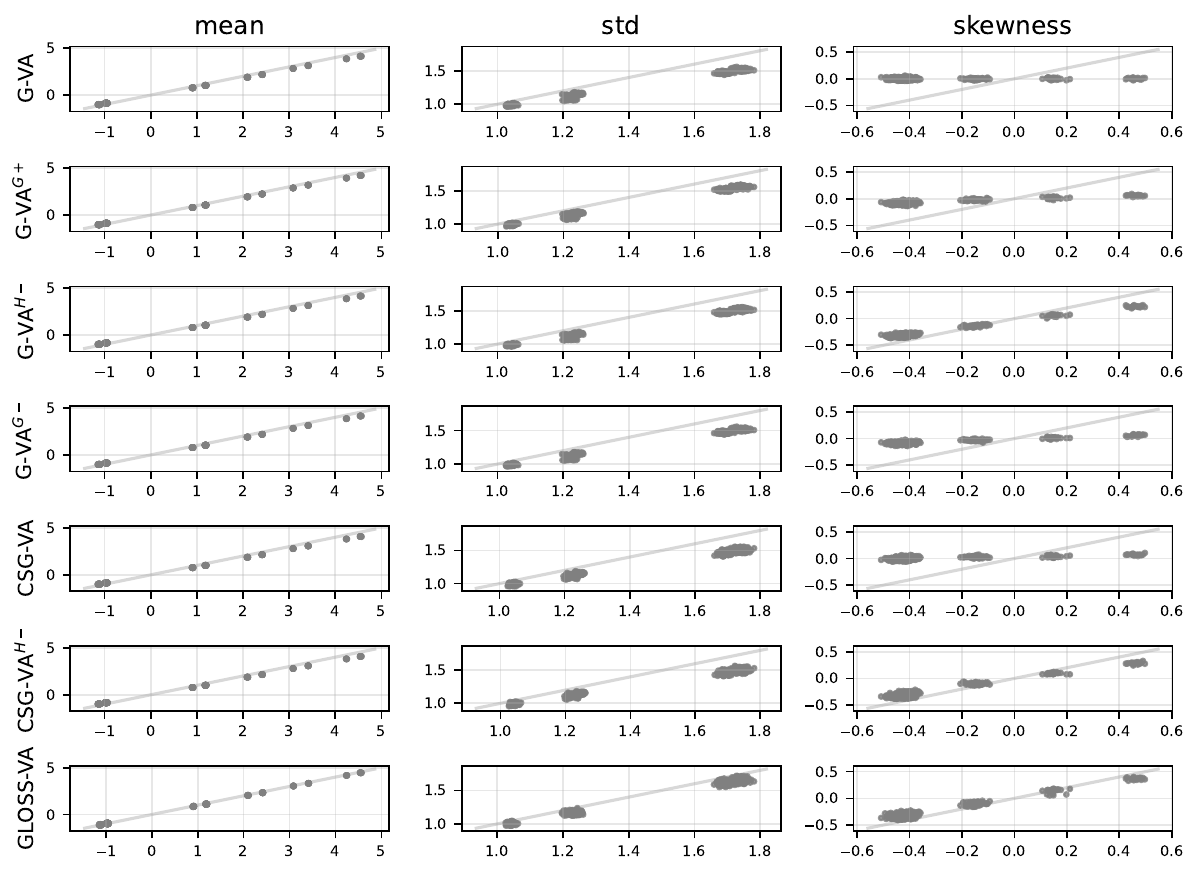}
    \caption{\small Logistic mixed model. Scatter plots show the mean (left), standard deviation (middle) and skewness (right) of the marginal posteriors of the random intercepts $\{b_i\}$ for each approximation method relative to estimates from MCMC. The dots will lie close to the diagonal if the approximation is accurate.}
    \label{fig:sixcities_local}
\end{figure}

 Figure~\ref{fig:sixcities_local} compares the mean, standard deviation and skewness of the marginal posteriors of the random effects computed using MCMC with each approximation method. While all methods are able to estimate the mean accurately, GLOSS-VA improves the estimation of both the skewness and standard deviation significantly, outperforming all other approximations. This drastic improvement in approximating the marginal posterior of the random effects, is likely the reason why the marginal posteriors of the global variables are approximated with much higher accuracy, since both $\beta_0$ and $\eta$ are closely linked to the random effects $b_1,\dots,b_n$ in the model formulation.

The good performance of CSG-VA$^{H-}$ demonstrates that our proposed conditional skewness correction can improve the base approximation, even when applied post-hoc. However, incorporating the skewness correction directly into the optimization yields even better results. In contrast, the post-hoc global skewness correction proposed by \citet{PozDurSza2024}, G-VA$^{G-}$, improves the base approximation G-VA only marginally in this example as the results from both approximations are virtually indistinguishable. G-VA$^{G+}$ slightly improves on G-VA$^{G-}$ illustrating that optimizing the skewness correction directly is helpful. However, the global skewness correction is inferior to the hierarchical correction, which is expected due to its increased flexibility.

\subsection{Poisson mixed model}
In the mixed effects Poisson regression model, a count response $y_{ij}$ is modeled using a Poisson distribution with mean $\mu_{ij}$, such that $$\log \mu_{ij}=x_{ij}^\top\beta+z_{ij}^\top b_i,$$
where $x_{ij}$ and $z_{ij}$ are covariates for the fixed and random effects respectively. Here, we consider a study on epileptic seizures for $n=59$ patients \citep{ThaVai1990} across $4$ visits. The response $y_{ij}$ denotes the number of epileptic seizures for the $i$th patient in the two weeks prior to visit $j$. The covariates include \texttt{trt} (indicates whether the patient is given the drug Progabide), \texttt{visit} (encodes the visit), \texttt{base} (log of 1/4 of the number of seizures experienced in the 8 weeks prior to treatment), and \texttt{age} (age of patient). We consider the linear predictor,
\begin{align*}
\log\mu_{ij}&=\beta_0 
+ \texttt{base}_i \beta_{\texttt{base}} 
+\texttt{trt}_i \beta_{\texttt{trt}} 
+\left(\texttt{base}_i\times\texttt{trt}_i\right) \beta_{\texttt{base}\times\texttt{trt}}
+\texttt{age}_i\beta_{\texttt{age}}
+\texttt{visit}_j\beta_{\texttt{visit}}
\\
& \quad + b_{i,0}+\texttt{visit}_j b_{\texttt{i,visit}}, 
\quad \text{for} \quad  i=1, \dots, 59, \;\;
j=1, \dots, 4,
\end{align*}
where $b_i=\left(b_{i,0},b_{i,\texttt{visit}}\right)^\top \sim\ND(0,CC^\top)$ is a patient-specific random effect.
Let $\beta = \left(\beta_0,\beta_{\texttt{base}}, \beta_{\texttt{trt}},\beta_{\texttt{base}\times\texttt{trt}},\beta_{\texttt{age}},\beta_{\texttt{visit}}\right)^\top$ be the fixed effect. The global variables are $\theta_G=\left(\beta^\top,\vech(C^*)^\top\right)^\top$. We consider the prior $\theta_G\sim\ND\left(0,10^2I_{9}\right)$. 

\begin{figure}[tb]
    \centering
    \includegraphics[width=0.95\linewidth,keepaspectratio]{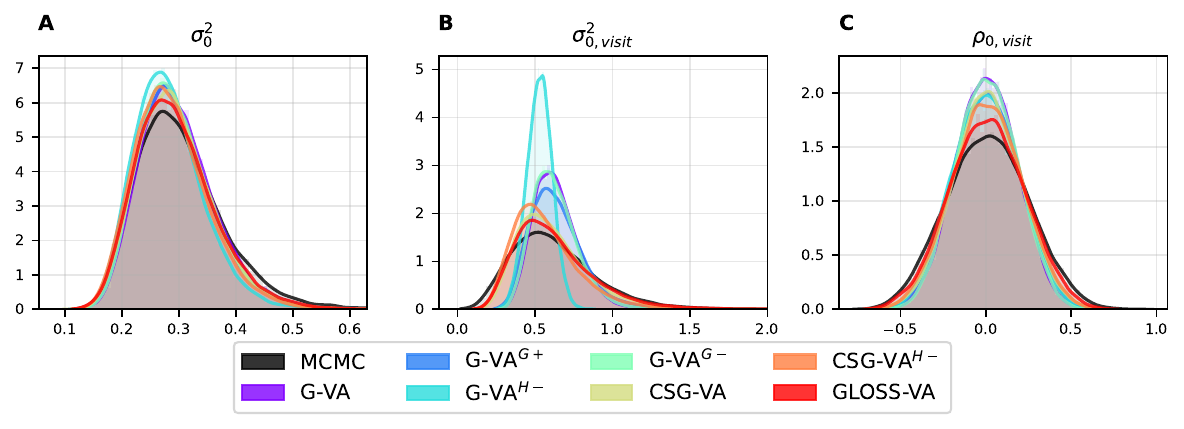}
    \caption{\small Poisson mixed model. Univariate marginal posteriors for the variances of the marginal random effects distribution (\textbf{A},\textbf{B}) and the correlation (\textbf{C}). 
    }
    \label{fig:epilepsy_sigma} 
\end{figure}

An investigation of the marginal posteriors of $\theta_G$, as well as the mean, standard deviation and skewness of the marginal posteriors of the random effects is given in Supporting Information~D.2. Similar observations to the previous application can be made, and GLOSS-VA shows marked improvements compared to all other approximation methods. Figure~\ref{fig:epilepsy_sigma} shows the marginal posteriors of the variances, $\sigma^2_0$ and $\sigma^2_{\text{visit}}$, of the random effects distribution as well as the correlation $\rho_{0,\text{visit}}$ between the two random effects, which are complex nonlinear combinations of $\vech(C^*)$. These parameters control the shape of the random effects distribution and can therefore be estimated well only if the posterior of $\theta_L$ is captured accurately. While all approximation methods can generally capture the shape of the true marginal posteriors, GLOSS-VA is closest to the results from MCMC.

\subsection{Discrete choice model}

Discrete choice models are widely used across various disciplines, and they provide valuable insights into the complex factors that drive individual decision-making. Here, we consider the  mixed multinomial logit model \citep[MMNL;][]{McfTra2000}, and apply it to a study on parking place choices \citep{IbeDelBorOrt2014}. In this study, $n=198$ respondents are faced with 8 choice scenarios, each having 3 alternatives:  free on-street parking (FSP), paid on-street parking (PSP), and paid parking in an underground car park (PUP). In each scenario, a different specification for the three variables, access time to parking (\texttt{at}), access time to destination (\texttt{td}), and parking fee (\texttt{fee}), is given. In addition, we consider the socio-economic binary variables, \texttt{li} (indicates if an individual has low income) and \texttt{res} (indicates if the respondent is a resident in the town).

Under the MMNL model, the probability that the $i$th individual $(i=1,\dots,n)$ chooses alternative $t\in\{\text{FSP},\text{PSP},\text{PUP}\}$ in the $j$th scenario $(j=1,\dots,8)$ is given by
\begin{equation*}
    \text{Pr}(y_{ij}=t)=\frac{\exp U_{ijt}}{\sum_{k\in\{\text{FSP, PSP, PUP}\}}\exp U_{ijk}},
\end{equation*}
where the individual and scenario specific utility is modeled as 
\begin{align*}
    U_{ijt} &= \texttt{at}_{jt}\beta_{\texttt{at}}+\texttt{td}_{jt}\beta_{\texttt{td}}+\texttt{fee}_{jt}\beta_{\texttt{fee}}+\left(\texttt{li}_i\times\texttt{fee}_{jt}\right)\beta_{\texttt{li}\times\texttt{fee}}\\
    & \quad +\left(\texttt{res}_i\times\texttt{fee}_{jt}\right)\beta_{\texttt{res}\times\texttt{fee}}+\texttt{at}_{jt}b_{i,\texttt{at}}+\texttt{td}_{jt}b_{i,\texttt{td}}+\texttt{fee}_{jt}b_{i,\texttt{fee}}.
\end{align*}
We consider the hierarchical prior \citep{HuaWan2013},
\begin{align*}
    \beta &\sim \ND(0,\sigma^2I), \\
    b_i=(b_{i,\texttt{at}},b_{i,\texttt{td}},b_{i,\texttt{fee}})^\top\mid\Sigma &\sim \ND(0,\Sigma), \\
    \Sigma\mid a_1,a_2,a_3  &\sim \IWD(\nu+2,2\nu\diag(a_1,a_2,a_3)), \\
    a_l &\sim\GD\left(1/2, 1/{A^2}\right), \qquad l=1,\dots,3,
\end{align*}
where $\sigma^2=10^6$, $\nu=2$, and $A=10^3$ are fixed hyperparameters, $\IWD(\nu,\Sigma)$ denotes the inverse Wishart distribution with $\nu$ degrees of freedom and scale matrix $\Sigma$, and $\GD(\alpha,\beta)$ denotes the Gamma distribution with shape $\alpha$ and rate $\beta$. The unconstrained global variables considered in our analysis are 
$$\theta_G=\left(\beta^\top,\vech(C^*)^\top,\log a_1,\log a_2,\log a_3\right)^\top,$$ 
where $C$ is the Cholesky factor of $\Omega=\Sigma^{-1}$. The induced prior under this reparameterization is derived in Supporting Information~C.

The total number of variables in this model is $\dim\theta=605$, and Bayesian inference via MCMC can already be challenging in medium sized MMNL due to poor mixing, and thus necessitating a large number of iterations \citep[e.g.~][]{HawHab2022}. Since MCMC sampling is often not feasible for the MMNL in practice, VI has emerged as a popular and effective alternative \citep[e.g.][]{Tan2017,Rod2022}. Here, we benchmark VI against the adaptive MCMC approach described in \citet{BanKruBieDazRas2020}. GLOSS-VA is approximately 33 times faster then MCMC on this data set. 

A visual comparison of the marginal posteriors of $\theta = (\theta_L, \theta_G)^\top$ estimated using MCMC and that obtained using each approximation method is given in Supporting Information~D.3. As in the previous examples, our novel conditional skewness correction is very helpful in approximating the complex posterior distribution, with CSG-VA$^{H-}$ and GLOSS-VA outperforming the other approximation methods. 

\begin{figure}[tb]
    \centering
    \includegraphics[width=0.95\linewidth,keepaspectratio]{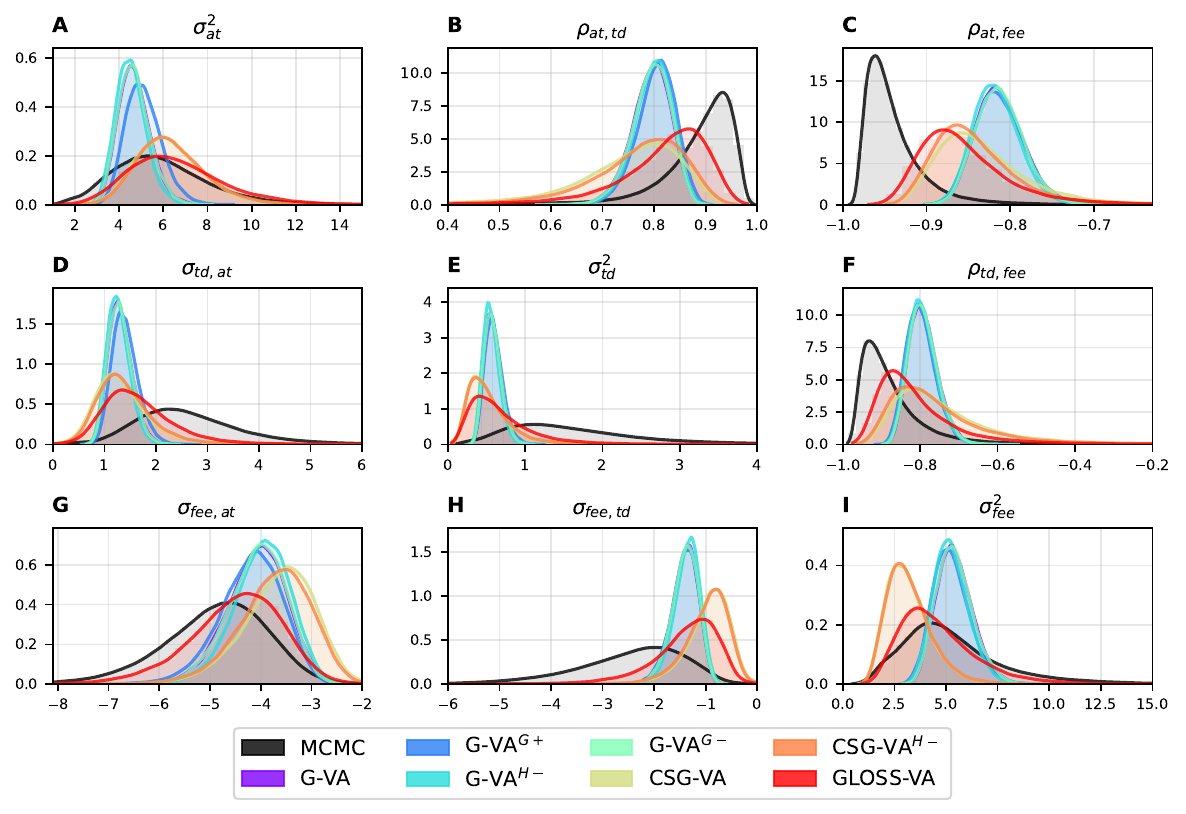}
    \caption{\small  MMNL model. Plots on the diagonal and lower triangular correspond to the respective entries of the covariance matrix $\Sigma$ of the random effects distribution. The diagonal represent marginal variances and the lower triangular represent bivariate covariances. Plots on the upper triangular report pairwise correlations, which are derived in closed form from entries of $\Sigma$. 
    }
    \label{fig:parking_sigma} 
\end{figure}

For the MMNL model, recovering $\Sigma$ is of particular interest as its entries are directly interpretable. \citet{IbeDelBorOrt2014} report strong heterogeneity for \texttt{at} and \texttt{fee}, which is expressed by a high variance in the marginals of the respective random effects distribution. There is a strong positive correlation between $b_{i,\texttt{at}}$ and $b_{i,\texttt{td}}$, while both $b_{i,\texttt{at}}$ and $b_{i,\texttt{td}}$ are negatively correlated with $b_{i,\texttt{fee}}$. Figure~\ref{fig:parking_sigma} summarizes the posterior of $\Sigma$, and shows the marginal posteriors for all variances, bivariate covariances and pairwise correlations. Recovering $\Sigma$ is challenging due to the complex structure of the model. A good variational approximation for $\Sigma$ necessitates not only a good estimation of the three-dimensional random effects, but also the posterior dependence structure of $p(\vech(C^*) \mid y)$. Again, the results of G-VA, G-VA$^{G-}$, and G-VA$^{G+}$ are virtually indistinguishable and they do not capture the shape of the true posterior very well. While both CSG-VA and CSG-VA$^{H-}$ yield better results than G-VA and G-VA$^{G-}$, the approximations from GLOSS-VA are closest to the density estimates from MCMC.

\subsection{Computational complexity}
To enable a fair comparison, all SGA algorithms are run using a conservative number of 150,000 iterations, and gradients are computed using automatic differentiation with \texttt{PyTorch}. The variational approximations exhibit a clear ordering in terms of their computational complexity. G-VA is the simplest approximation, with the lowest number of parameters.  As G-VA$^{G-}$ is a perturbation of the fitted G-VA and sampling from the skew-symmetric distribution is fast, its computational complexity is similar to G-VA. CSG-VA introduces additional parameters to accommodate a more flexible variational family, which leads to a slight increase in computational cost compared to G-VA. Similarly, CSG-VA$^{H-}$ is a post-hoc perturbation of CSG-VA and thus incurs a comparable cost to CSG-VA. Although GLOSS-VA and CSG-VA have the same number of parameters, the inclusion of the skewness correction in the optimization process increases the computational complexity. Estimating the ELBO for GLOSS-VA, as described in Section \ref{sec:optimization}, also requires evaluating a significant number of additional terms. 

Computation times for all variational methods and the estimated ELBO during training are shown in Supporting Information~D.4. GLOSS-VA leads to the highest ELBO after training for all examples considered and convergence is reached well before the maximum 150,000 iterations.

\section{Summary and discussion}\label{sec:discussion}
In this article, we have introduced a novel approach to improve Gaussian variational approximations by incorporating skewness corrections tailored for hierarchical models with both global and local parameters. Our method builds upon the skewness correction framework proposed by \cite{PozDurSza2024}, and extends its applicability to conditionally structured Gaussian approximations. This allows for enhanced accuracy in capturing posterior distributions that exhibit skewness, a limitation often encountered with traditional Gaussian variational approaches.

Our contributions are threefold. First, we successfully implemented skewness corrections in hierarchical models, utilizing a decomposition strategy that separately addresses global and local parameters with adaptive skew-symmetric adjustments. This provides a more accurate representation of complex posterior distributions in such models. Secondly, we developed an efficient technique for optimization of the novel variational family using a variation of the reparametrization trick. Thirdly, we demonstrate superior performance in terms of posterior approximation accuracy compared to existing methods in various examples.

Future research can integrate our approach with other advanced variational techniques such as amortization and richer variational families, to enhance accuracy in more complex settings. In addition, further theoretical exploration of the asymptotic properties of our corrected variational approximations in large-scale hierarchical models can provide deeper insights and validation of the method's robustness across different contexts, and potentially lead to further improved variational families. In this work, we focus on models in which the local parameters are mutually independent given the global parameters. Extending our approach to accommodate models with more complex dependency structures, such as state space models or generalized regression models with crossed random effects, presents a promising direction for future research that we are currently pursuing.

\FloatBarrier
\bibliography{bib}
\FloatBarrier
\appendix
\newpage 

\section{Derivation of the ELBO}
Let $\widetilde{b_i}(\varepsilon_i,\theta_G)=\mu_i(\theta_G)+T_i(\theta_G)^{-\top}\varepsilon_i$, $\widetilde{\theta_G}(\varepsilon_G)=\mu_G+T_G^{-\top}\varepsilon_G$ and $h_G(\theta_G) = p(\theta_G)$. Then,
\begin{align*}
    &\mathbb{E}_{q_\lambda(\theta)}[\log(p(y,\theta))]=\int \left[\log\left(h_G(\theta_G)\right)+\sum_{i=1}^n\log\left(h_i(b_i\mid\theta_G)\right)\right]q_\lambda(\theta) d\theta\\
    &=\int\indicator{u_G\leq w(\widetilde{\theta_G}(\varepsilon_G))}\Bigg\{\log\left(h_G(\widetilde{\theta_G}(\varepsilon_G))\right)+\sum_{i=1}^n\bigg[\indicator{u_i\leq w(\widetilde{b_i}(\varepsilon_i,\widetilde{\theta_G}(\varepsilon_G)))}
    \\
    & \times \log\left(h_i(\widetilde{b_i}(\varepsilon_i,\widetilde{\theta_G}(\varepsilon_G))\mid\widetilde{\theta_G}(\varepsilon_G))\right) +\indicator{u_i> w(\widetilde{b_i}(\varepsilon_i,\widetilde{\theta_G}(\varepsilon_G)))} 
    \\
    & \times \log\left(h_i(2\mu_i(\widetilde{\theta_G}(\varepsilon_G))-\widetilde{b_i}(\varepsilon_i,\widetilde{\theta_G}(\varepsilon_G))\mid\widetilde{\theta_G}(\varepsilon_G))\right)\bigg]\Bigg\}
    \\&+\indicator{u_G>w(\widetilde{\theta_G}(\varepsilon_G))}\Bigg\{\log\left(h_G(2\mu_G-\widetilde{\theta_G}(\varepsilon_G))\right)+\sum_{i=1}^n\bigg[\indicator{u_i\leq w(\widetilde{b_i}(\varepsilon_i,2\mu_G-\widetilde{\theta_G}(\varepsilon_G)))}\\&\times\log\left(h_i(\widetilde{b_i}(\varepsilon_i,2\mu_G-\widetilde{\theta_G}(\varepsilon_G))\mid2\mu_G-\widetilde{\theta_G}(\varepsilon_G))\right)+\indicator{u_i> w(\widetilde{b_i}(\varepsilon_i,2\mu_G-\widetilde{\theta_G}(\varepsilon_G)))}\\&\times\log\left(h_i(2\mu_i(2\mu_G-\widetilde{\theta_G}(\varepsilon_G))-\widetilde{b_i}(\varepsilon_i,2\mu_G-\widetilde{\theta_G}(\varepsilon_G))\mid2\mu_G-\widetilde{\theta_G}(\varepsilon_G))\right)\bigg]\Bigg\}dp(u,\varepsilon).
\end{align*}
Integration over $u$ yields
\begin{align*}
    \int& w(\widetilde{\theta_G}(\varepsilon_G))\Bigg\{\log\left(h_G(\widetilde{\theta_G}(\varepsilon_G))\right)+\sum_{i=1}^n\bigg[w(\widetilde{b_i}(\varepsilon_i,\widetilde{\theta_G}(\varepsilon_G)))\log\left(h_i(\widetilde{b_i}(\varepsilon_i,\widetilde{\theta_G}(\varepsilon_G))\mid\widetilde{\theta_G}(\varepsilon_G))\right)\\&+(1-w(\widetilde{b_i}(\varepsilon_i,\widetilde{\theta_G}(\varepsilon_G))))\log\left(h_i(2\mu_i(\widetilde{\theta_G}(\varepsilon_G))-\widetilde{b_i}(\varepsilon_i,\widetilde{\theta_G}(\varepsilon_G))\mid\widetilde{\theta_G}(\varepsilon_G))\right)\bigg]\Bigg\}
    \\&+\left(1-w(\widetilde{\theta_G}(\varepsilon_G))\right)\Bigg\{\log\left(h_G(2\mu_G-\widetilde{\theta_G}(\varepsilon_G))\right)+\sum_{i=1}^n\bigg[w(\widetilde{b_i}(\varepsilon_i,2\mu_G-\widetilde{\theta_G}(\varepsilon_G)))\\&\times\log\left(h_i(\widetilde{b_i}(\varepsilon_i,2\mu_G-\widetilde{\theta_G}(\varepsilon_G))\mid2\mu_G-\widetilde{\theta_G}(\varepsilon_G))\right)+(1-w(\widetilde{b_i}(\varepsilon_i,2\mu_G-\widetilde{\theta_G}(\varepsilon_G))))\\&\times\log\left(h_i(2\mu_i(2\mu_G-\widetilde{\theta_G}(\varepsilon_G))-\widetilde{b_i}(\varepsilon_i,2\mu_G-\widetilde{\theta_G}(\varepsilon_G))\mid2\mu_G-\widetilde{\theta_G}(\varepsilon_G))\right)\bigg]\Bigg\}dp(\varepsilon).
\end{align*}
The same strategy can be applied to the entropy term.
\begin{align*}
    &\mathbb{E}_{ q_\lambda(\theta)} \left[ \log q_\lambda(\theta) \right] =
    (n+1)\log 2 +\int \bigg\{ \log \varphi(\theta_G;\mu_G,\Sigma_G) + \log w(\theta_G) 
    \\
    &  + \sum_{i=1}^n \left[\log \varphi(b_i;\mu_i(\theta_G),\Sigma_i(\theta_G)) + \log w(b_i)\right] \bigg\} q_\lambda(\theta) d\theta
    \\
    &= C+ \int \bigg\{ \sum_{j=1}^{d_G}\left(T_G^*\right)_{jj}+\norm{T_G^\top(\theta_G-\mu_G)}_2^2+ \log w(\theta_G) \\& + \sum_{i=1}^n\bigg[ \sum_{j=1}^{d_i} \left(T_i(\theta_G)^*\right)_{jj}+\norm{T_i(\theta_G)^\top(b_i-\mu_i(\theta_G))}_2^2+\log w(b_i)\bigg] \bigg\} q_\lambda(\theta) d\theta, 
\end{align*}
which becomes
\begin{align*}
    &\mathbb{E}_{ q_\lambda(\theta)}\left[\log q_\lambda(\theta)\right] 
    = C + \sum_{j=1}^{d_G}\left(T_G^*\right)_{jj} + \int  \norm{T_G^\top(\widetilde{\theta_G}(\varepsilon_G)-\mu_G)}_2^2 + w(\widetilde{\theta_G}(\varepsilon_G))\bigg[\log w(\widetilde{\theta_G}(\varepsilon_G))
    \\
    &+\sum_{i=1}^n\sum_{j=1}^{d_i} \left(T_i(\widetilde{\theta_G}(\varepsilon_G))^*\right)_{jj} 
    + \sum_{i=1}^n \norm{T_i(\widetilde{\theta_G}(\varepsilon_G))^\top(\widetilde{b_i}(\varepsilon_i,\widetilde{\theta_G}(\varepsilon_G))-\mu_i(\widetilde{\theta_G}(\varepsilon_G)))}_2^2
    \\
    &+\sum_{i=1}^n w(\widetilde{b_i}(\varepsilon_i,\widetilde{\theta_G}(\varepsilon_G))) \log w(\widetilde{b_i}(\varepsilon_i,\widetilde{\theta_G}(\varepsilon_G))) 
    + \left(1-w(\widetilde{b_i}(\varepsilon_i,\widetilde{\theta_G}(\varepsilon_G)))\right) 
     \\
     &\times \log\left(1-w(\widetilde{b_i}(\varepsilon_i,\widetilde{\theta_G}(\varepsilon_G)))\right) \bigg]
    +\left(1-w(\widetilde{\theta_G}(\varepsilon_G))\right)\bigg[\log \left(1-w(\widetilde{\theta_G}(\varepsilon_G))\right)
    \\
    &+\sum_{i=1}^n\sum_{j=1}^{d_i} \left(T_i(2\mu_G-\widetilde{\theta_G}(\varepsilon_G))^*\right)_{jj} 
    \\
    &+ \sum_{i=1}^n \norm{T_i(2\mu_G-\widetilde{\theta_G}(\varepsilon_G))^\top(\widetilde{b_i}(\varepsilon_i,2\mu_G-\widetilde{\theta_G}(\varepsilon_G))-\mu_i(2\mu_G-\widetilde{\theta_G}(\varepsilon_G)))}_2^2
    \\
    &+\sum_{i=1}^n w(\widetilde{b_i}(\varepsilon_i,2\mu_G-\widetilde{\theta_G}(\varepsilon_G))) \log w(\widetilde{b_i}(\varepsilon_i, 2\mu_G-\widetilde{\theta_G}(\varepsilon_G))) 
    \\
    &+ \left(1-w(\widetilde{b_i}(\varepsilon_i,2\mu_G-\widetilde{\theta_G}(\varepsilon_G)))\right) \log\left(1-w(\widetilde{b_i}(\varepsilon_i,2\mu_G-\widetilde{\theta_G}(\varepsilon_G)))\right) \bigg] dp(\varepsilon).
\end{align*}
Note that $C$ is a constant independent of $\lambda$.

\section{Discussion on the global skewness approximation}

Following \citet{PozDurSza2024}, the optimal skewness correction for the global parameters $\theta_G$ is given as 
\begin{equation*}
    w(\theta_G)=\frac{h(\theta_G)}{h(\theta_G)+h(2\mu_G-\theta_G)},
\end{equation*}
where $h(\theta_G)=p(\theta_G)p(y\mid\theta_G)$ is the kernel of the marginal posterior $p(\theta_G\mid y)$. Since $p(\theta_G\mid y)$ is usually intractable, our approach uses an approximation $\Tilde{h}(\theta_G)$ to $h(\theta_G)$ and we will now discuss the influence of this approximation. 

Note that 
$$h(\theta_G)=\Tilde{h}(\theta_G)\frac{q_\lambda(\mu(\theta_G)\mid\theta_G)}{p(\mu(\theta_G)\mid \theta_G,y)},$$ where $\mu(\theta_G)=(\mu_1(\theta_G)^\top,\dots,\mu_n(\theta_G)^\top)^\top$.
This shows that the approximation is tight if $q_\lambda(\mu(\theta_G)\mid\theta_G)\approx p(\mu(\theta_G)\mid\theta_G,y)$, that is $q_\lambda(\theta_L\mid\theta_G)$ is a good approximation to $p(\mu(\theta_G)\mid\theta_G,y)$ at its mean $\mu(\theta_G)$. 
In the special case that $p(b_i,\theta_G\mid y)$ is jointly Gaussian for all $i=1,\dots,n$, $q_\lambda(\theta_L\mid\theta_G)=p(\theta_L\mid\theta_G,y)$ and thus $\Tilde{h}(\theta_G)=h(\theta_G)$. 

Our method of approximating the marginal posterior is also closely linked to the commonly applied Laplace approximation, which is given by matching the posterior mode and the corresponding Hessian to derive a Gaussian approximation of $p(b_i\mid\theta_G,y_i), i=1,\dots,n$. \citet{durante+ps23} derive a skewed Bernstein-von Mises therorem that considers a skew symmetric normal distribution with faster rate of convergence to the posterior than the regular Bernstein-von Mises theorem. This supports the use of the skewed approximation $q_\lambda(\theta_L\mid\theta_G)$ in our context as offering a more reliable approximation than a Gaussian approximation.

In summary, the quality of our approximation to the skewness correction
for $\theta_G$ is directly controlled by the quality of the variational approximation for $\theta_L\mid\theta_G$.

\section{Reparameterization of MMNL model}
Under the MMNL model \citep{McfTra2000}, the probability that the $i$th individual ($i=1,\dots,n$) chooses alternative $t$ in scenario $j$ is given as \begin{equation*}
    \text{Pr}(y_{ij}=t)=\frac{\exp U_{ijt}}{\sum_{k}\exp U_{ijk}}.
\end{equation*}
We model the individual and scenario specific utility as 
\begin{align*}
    U_{ijt} = x_{jt}^\top\beta+z_{jt}^\top b_i,
\end{align*}
where $x_{jt}$ and $z_{jt}$ are the scenario and alternative specific vectors of covariates, and $\beta\in\mathbb{R}^{d_G}$ and $b\in\mathbb{R}^{d_L}$ denote the fixed and random effects respectively. We consider the hierarchical prior,
\begin{align*}
    \beta &\sim \ND(0,\sigma^2I), \\
    b_i\vert\Sigma &\sim \ND(0,\Sigma), \\
    \Sigma\vert a_1,\dots,a_{d_L}  &\sim \IWD(\nu+d_L-1,2\nu\diag(a_1,\dots,a_{d_L})), \\
    a_l &\sim\GD\left(\frac{1}{2},\frac{1}{A^2}\right), \qquad l=1,\dots,d_L.
\end{align*}
where $\sigma^2=10^6$, $\nu=2$, and $A=10^3$ are fixed hyperparameters. Let $\Omega=\Sigma^{-1}=CC^\top$, where $C$ is a lower triangular matrix with positive diagonal. Then $$\theta_G=\left(\beta^\top,\vech(C^*)^\top,\log a_1,\dots,\log a_{d_L}\right)^\top$$ denotes the unconstrained  global variables. 

We have $\Omega\vert a_1,\dots,a_{d_L} \sim \WD\left(\nu+d_L-1,\diag\left(\frac{1}{2\nu a_1},\dots,\frac{1}{2\nu a_{d_L}}\right)\right)$. In addition,
\begin{equation*}
    \det \left( \nabla_{\vech(C^*)} \vech(\Omega) \right) 
    = \det \left( L\left(I_{d_L^2}+K\right)\left(C\otimes I_{d_L} \right) L^\top \right) \prod_{l=1}^{d_L} C_{ll},
\end{equation*}
where $L$ is the elimination matrix, so that $\vech(A)=L\vect(A)$ for all $A\in\mathbb{R}^{d_L\times d_L}$, and $K$ is the commutation matrix, so that $\vect(A)=K\vect(A^\top)$ for all $A\in\mathbb{R}^{d_L\times d_L}$ \citep[e.g.][]{QuiNotKoh2023}. Finally, 
\begin{equation*}
    \nabla_{\log a_l} a_l = a_l,\quad \text{for} \quad l=1,\dots,d_L.
\end{equation*}
Hence $p(\theta_G)=p(\beta)p(\Omega\vert a_1,\dots,a_{d_L})\det\left(L\left(I_{d_L^2}+K\right)\left(C\otimes I_{d_L}\right)L^\top\right)\prod_{l=1}^{d_L} \left[C_{ll} a_l p(a_l)\right]$ is the prior for the transformed $\theta_G$.

\section{Additional results for applications}
\subsection{Logistic mixed model}
\begin{figure}[htb!]
    \centering
    \includegraphics[width=0.9\linewidth,keepaspectratio]{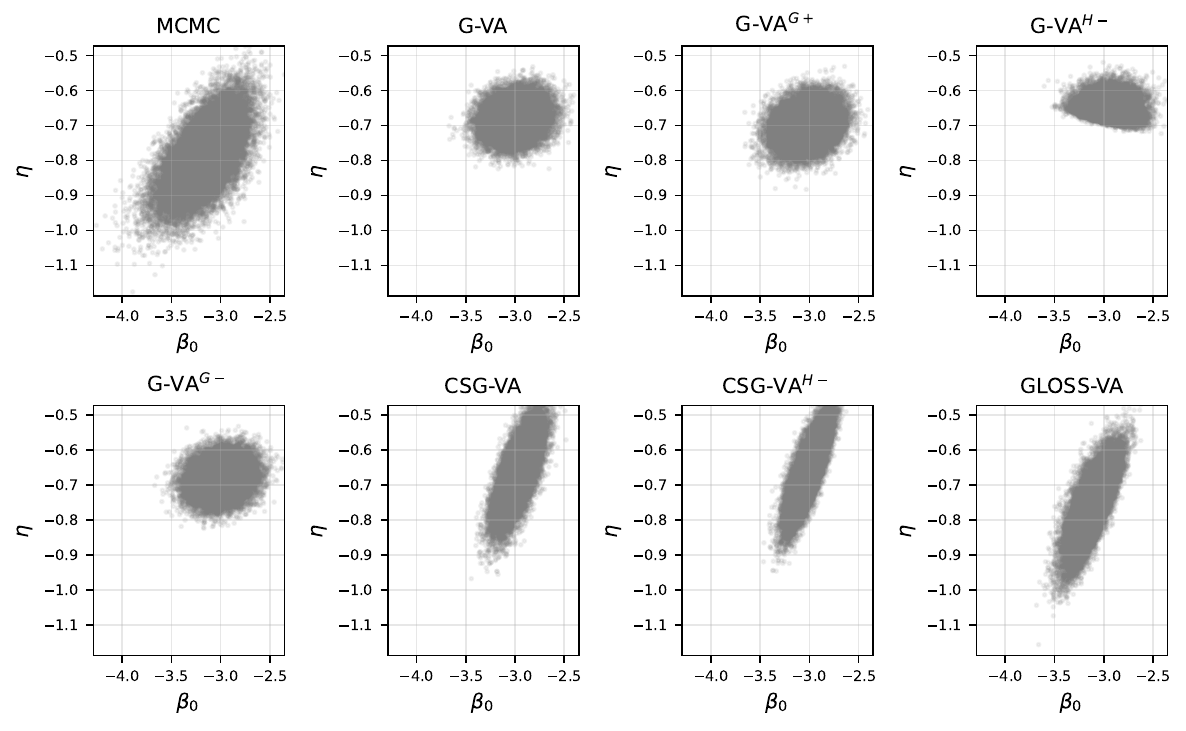}
    \caption{\small Logistic mixed model. Scatter plots display the samples generated from $p(\beta_0,\eta\mid y)$ using MCMC and each approximation method.}
    \label{app:fig:sixcities_bivariate}
\end{figure}

Figure~\ref{app:fig:sixcities_bivariate} shows scatter plots of the samples generated from $p(\beta_0,\eta\mid y)$ using MCMC and each approximation method, for the random intercept logistic regression example as presented in the main paper.

\begin{figure}[tb]
    \centering
    \includegraphics[width=0.9\linewidth,keepaspectratio]{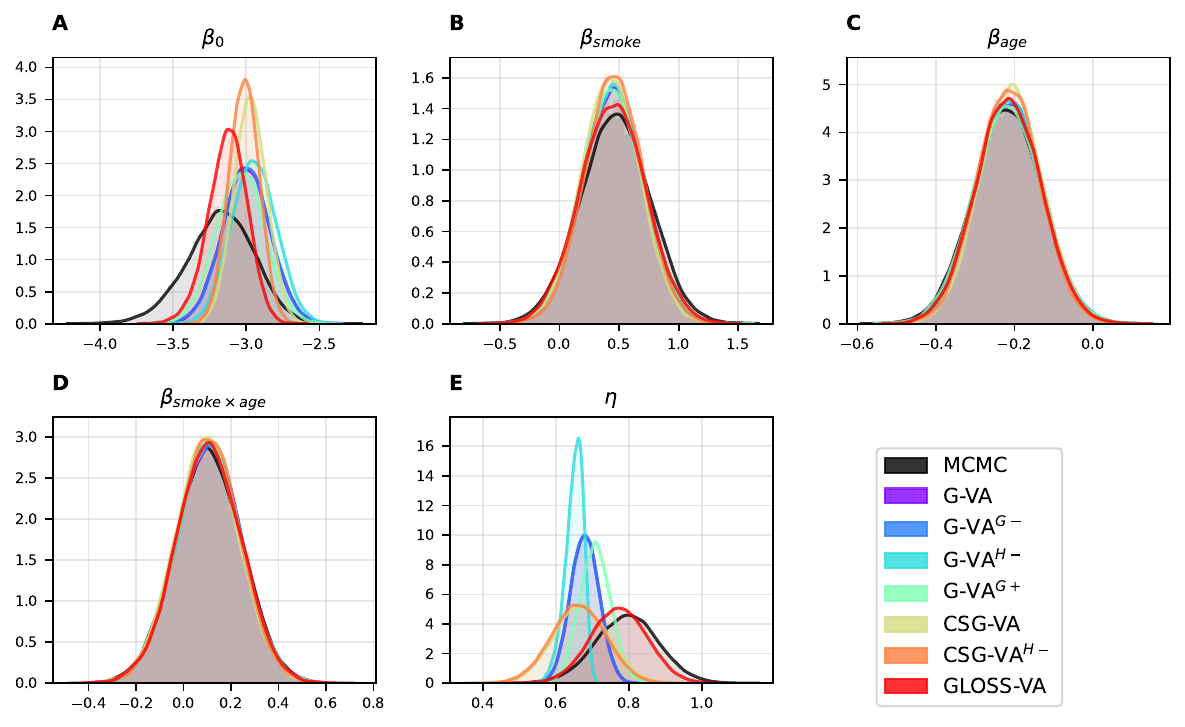}
    \caption{\small  Logistic mixed model with Huang-Wang prior. Univariate marginal posteriors for all global parameters. 
    }
    \label{fig:sixcities_alt_global}
\end{figure}

\begin{figure}[tb]
    \centering
    \includegraphics[width=0.95\linewidth,keepaspectratio]{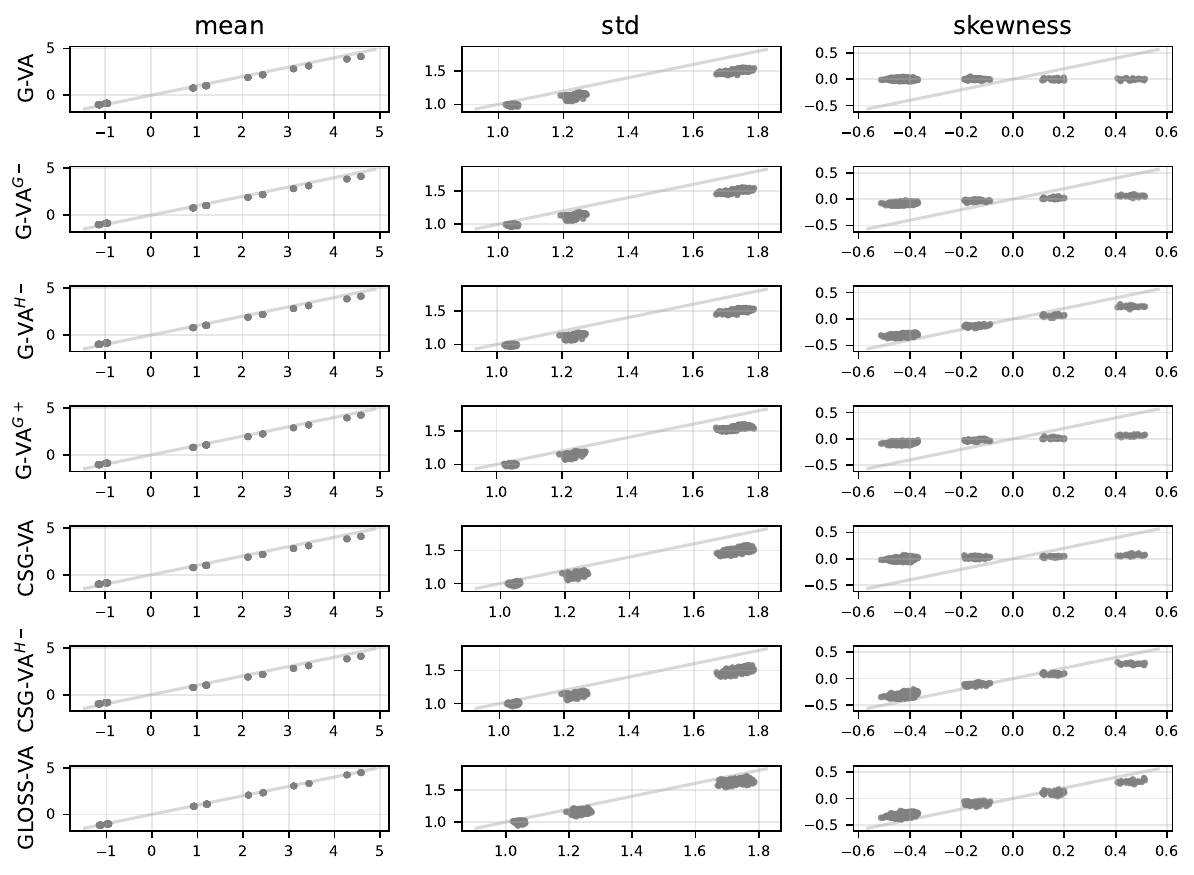}
    \caption{\small Logistic mixed model with Huang-Wang prior. Scatter plots show the mean (left), standard deviation (middle) and skewness (right) of the marginal posteriors of the random intercepts $\{b_i\}$ for each approximation method relative to estimates from MCMC. The dots will lie close to the diagonal if the approximation is accurate.}
    \label{fig:sixcities_alt_local}
\end{figure}

To analyze how sensitive our approach is to different prior structures, we consider a different parameterization of the model using the Huang-Wand prior \citep{HuaWan2013} on the prior standard deviation of $b_i$. Again we consider the random intercept logistic regression model,
\begin{equation*}
\log\left(\frac{\text{Pr}(y_{ij}=1)}{1-\text{Pr}(y_{ij}=1)}\right)=\beta_{0}+\texttt{smoke}_{ij}\beta_{\texttt{smoke}}+\texttt{age}_{ij}\beta_{\texttt{age}}+\left(\texttt{smoke}_{ij}\times\texttt{age}_{ij}\right)\beta_{\texttt{smoke}\times\texttt{age}}+b_i,
\end{equation*}
where $\beta=\left(\beta_{0},\beta_{\texttt{smoke}},\beta_{\texttt{age}},\beta_{\texttt{smoke}\times\texttt{age}}\right)^\top$ is the vector of fixed effects. In difference to the analysis shown in Section~5.1 of the main paper,  we use the prior
\begin{align*}
    \beta&\sim\ND(0,10^2I_5)\\
    b_i&\sim\ND(0,\sigma^2), i=1,\dots,n\\
    \sigma&\sim\HalfTD(2,10),
\end{align*}
where $\HalfTD(\nu,s)$ denotes a half-t distribution with $\nu$ degrees of freedom and scale parameter $s$. Writing $\eta=\log\sigma$, the vector of unrestricted global variables is $\theta_G=\left(\beta^\top,\eta\right)^\top$. Figures~\ref{fig:sixcities_alt_global} and \ref{fig:sixcities_alt_local} illustrate the performance of the VI benchmarks on $\theta_G$ and $\theta_L$ under this model respectively. We can draw similar conclusions to the analysis presented in Section~5.1 and the results are insensitive to this prior choice. In particular, GLOSS-VA outperforms the other benchmarks in recovering the marginal posterior for the prior standard deviation of $b_i$.

\subsection{Poisson mixed model}

\begin{figure}[htb!]
    \centering
    \includegraphics[width=0.9\linewidth,keepaspectratio]{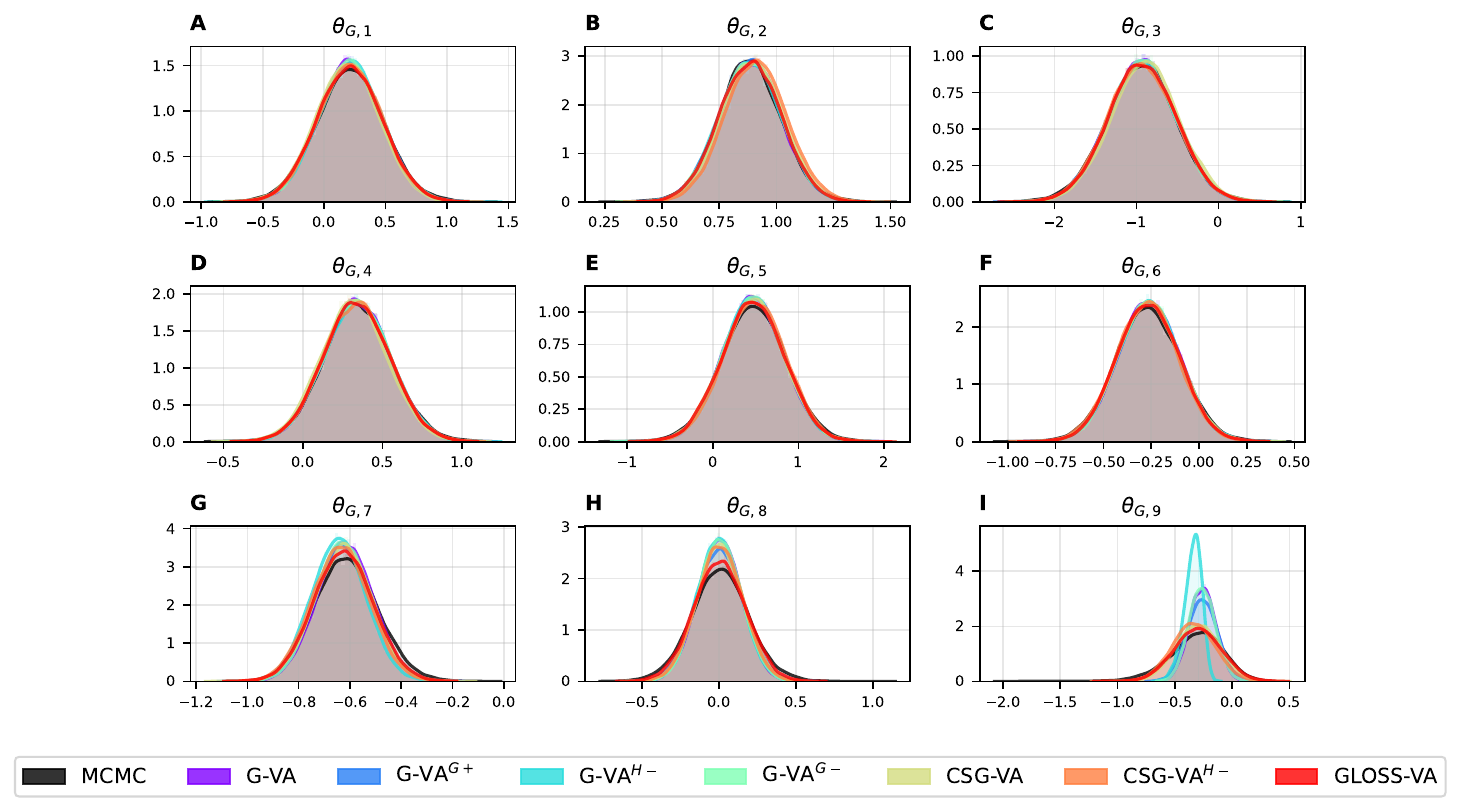}
    \caption{\small Poisson mixed model. Marginal posteriors for $\theta_G$.}
    \label{app:fig:epilepsy_global}
\end{figure}

\begin{figure}[tbh]
    \centering
    \includegraphics[width=0.95\linewidth,keepaspectratio]{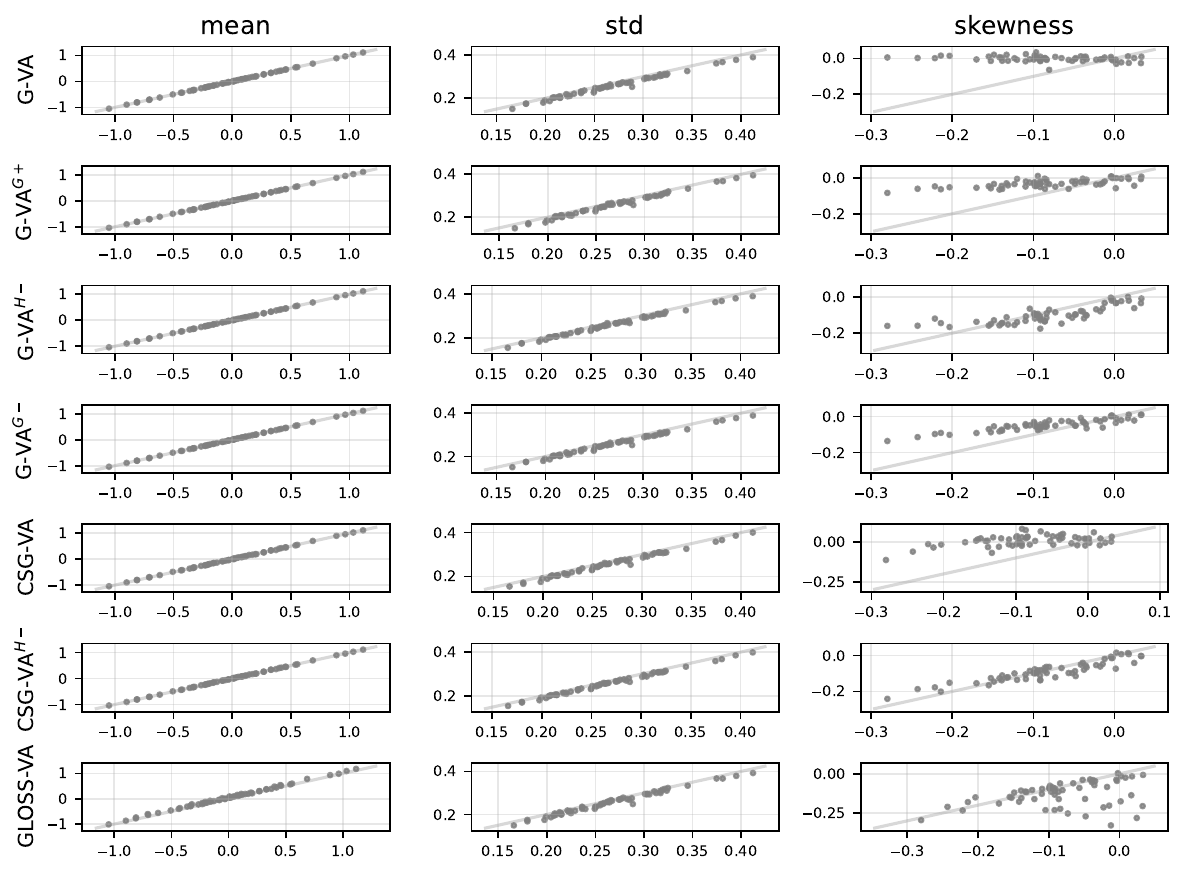}
    \includegraphics[width=0.95\linewidth,keepaspectratio]{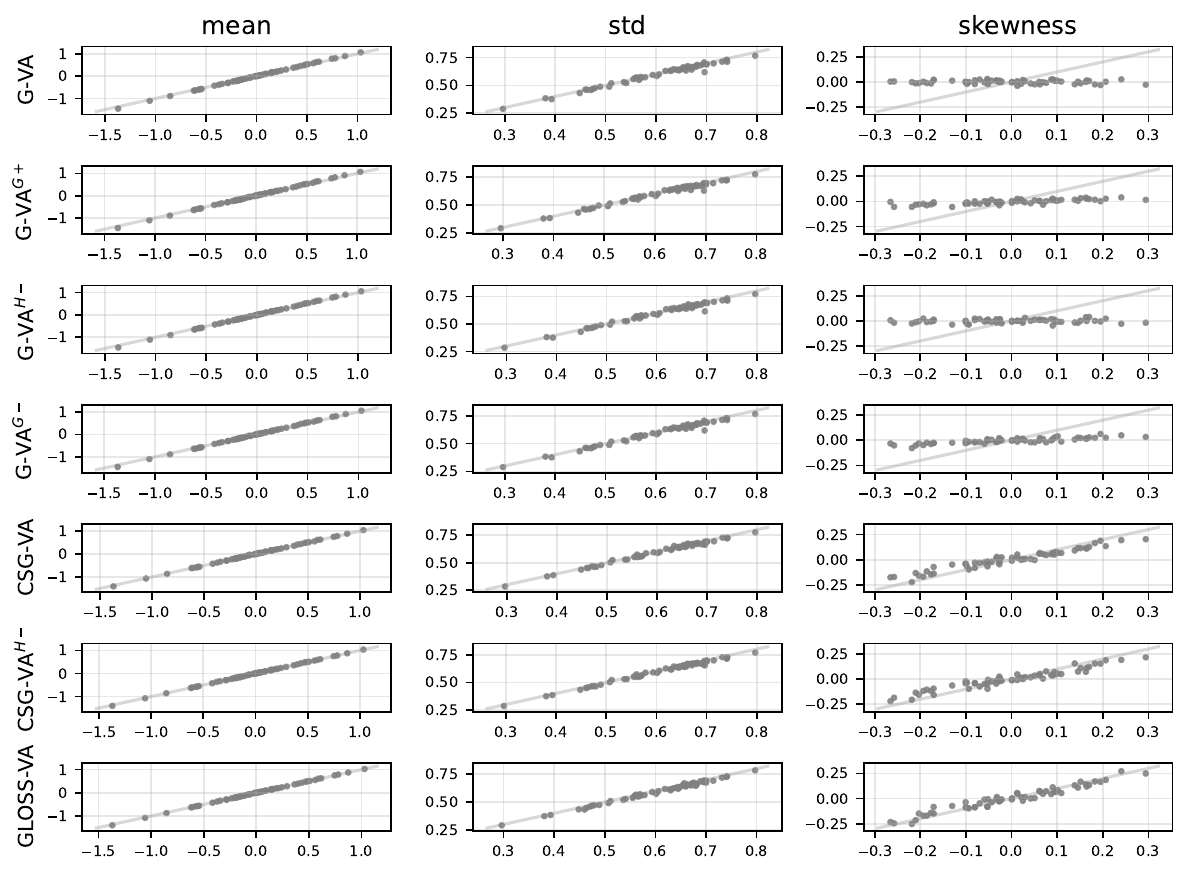}
    \caption{\small  Poisson mixed model. Scatter plots for the mean (left), standard deviation (middle) and skewness (right) for the marginal posteriors of the random intercepts $\{b_i\}$ for each approximation method (columns) relative to estimates from MCMC for $b_{i,0}$ (top) and $b_{i,\texttt{visit}}$ (bottom).}
    \label{app:fig:epilepsy_local}
\end{figure}

Figure~\ref{app:fig:epilepsy_global} and Figure~\ref{app:fig:epilepsy_local} summarize the marginal posterior approximations of the global and latent variables for the Epilepsy data discussed in Section 3.2 respectively. 

\subsection{Discrete choice model}
\begin{figure}[tbh]
    \centering
    \includegraphics[width=0.95\linewidth,keepaspectratio]{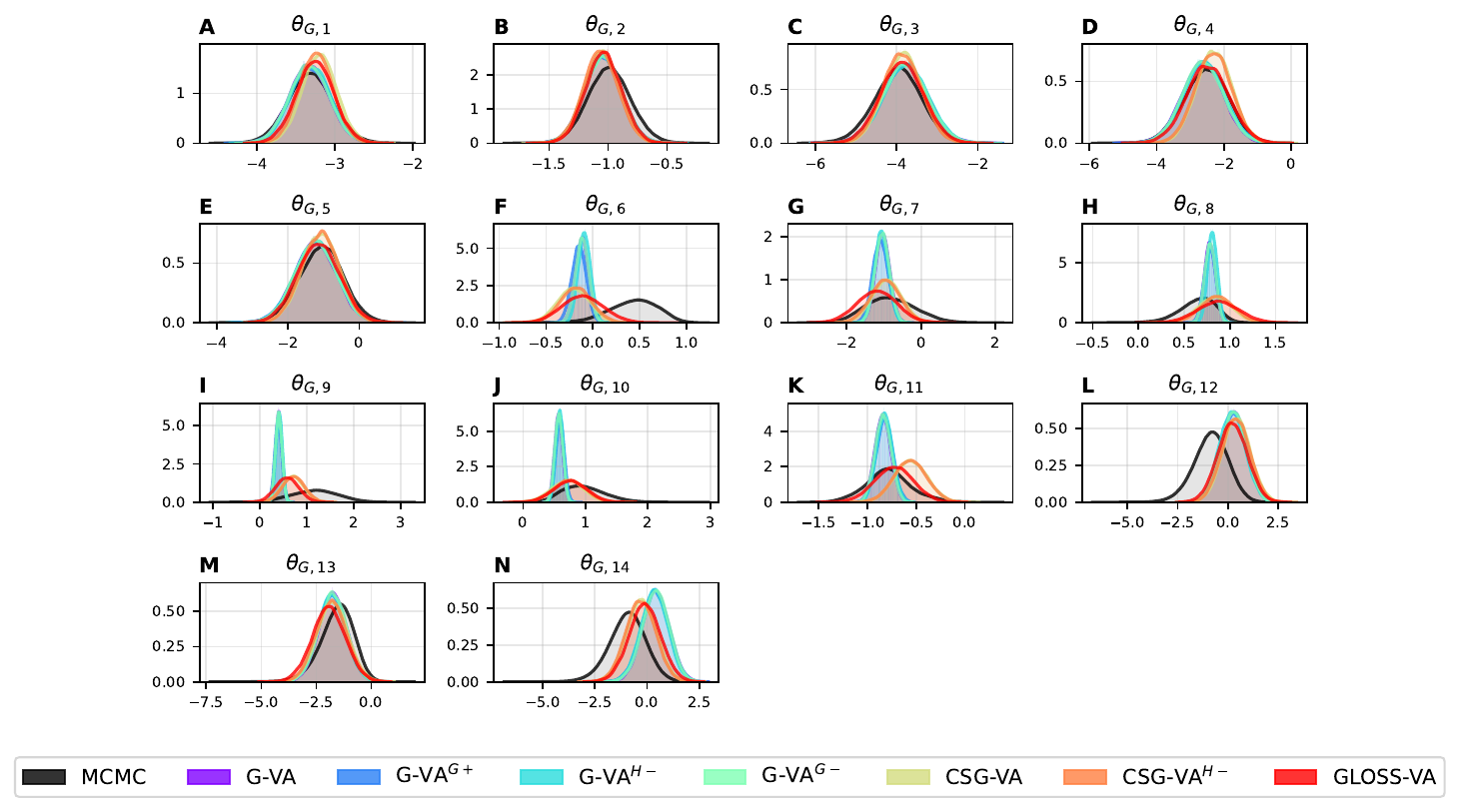}
    \caption{\small  MMNL model. Marginal posteriors for $\theta_G$.}
    \label{app:fig:parking_global}
\end{figure}

\begin{figure}[tbh]
    \centering
    \includegraphics[width=0.95\linewidth,keepaspectratio]{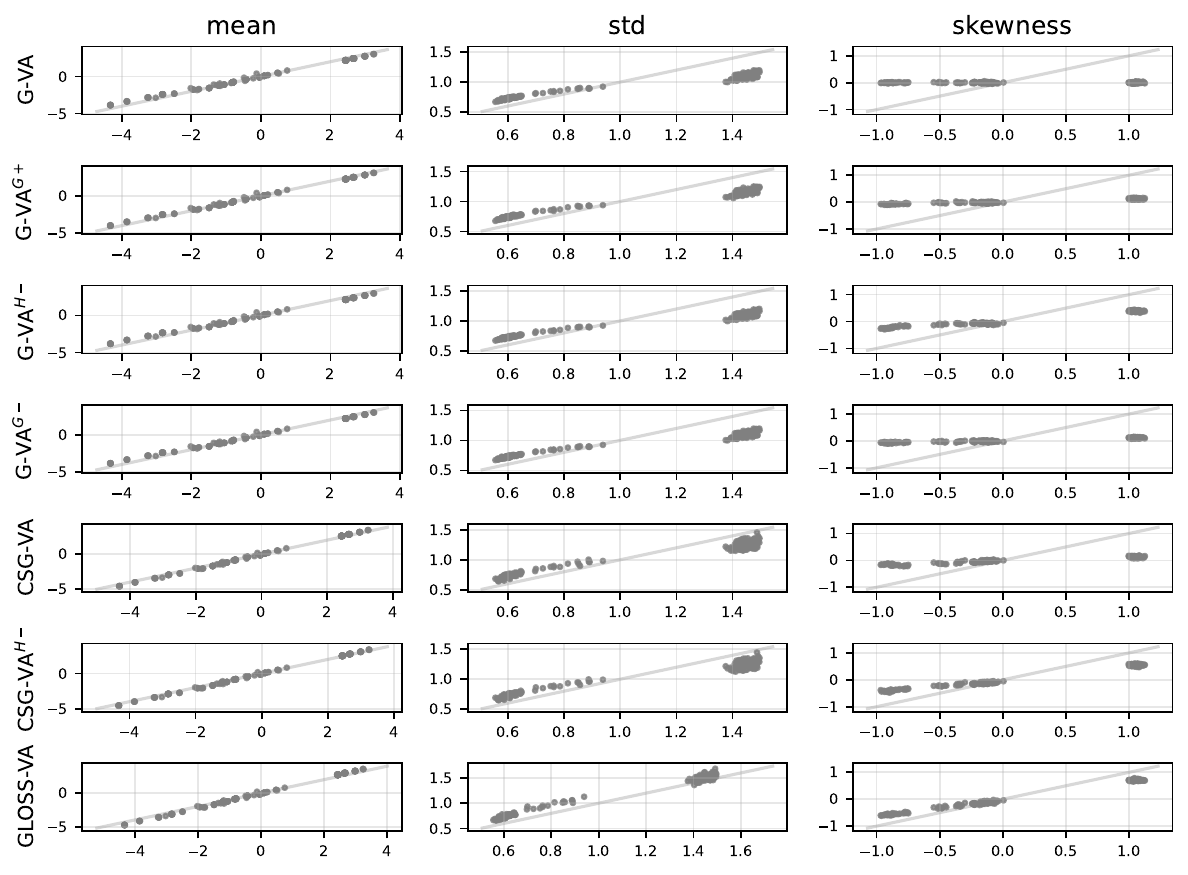}
    \caption{\small  MMNL model. Scatter plots for the mean (left), standard deviation (middle) and skewness (right) for the marginal posteriors of $b_{i,\texttt{at}}$ for each approximation method (columns) relative to estimates from MCMC.}
    \label{app:fig:parking_local1}
\end{figure}

\begin{figure}[tbh]
    \centering
    \includegraphics[width=0.95\linewidth,keepaspectratio]{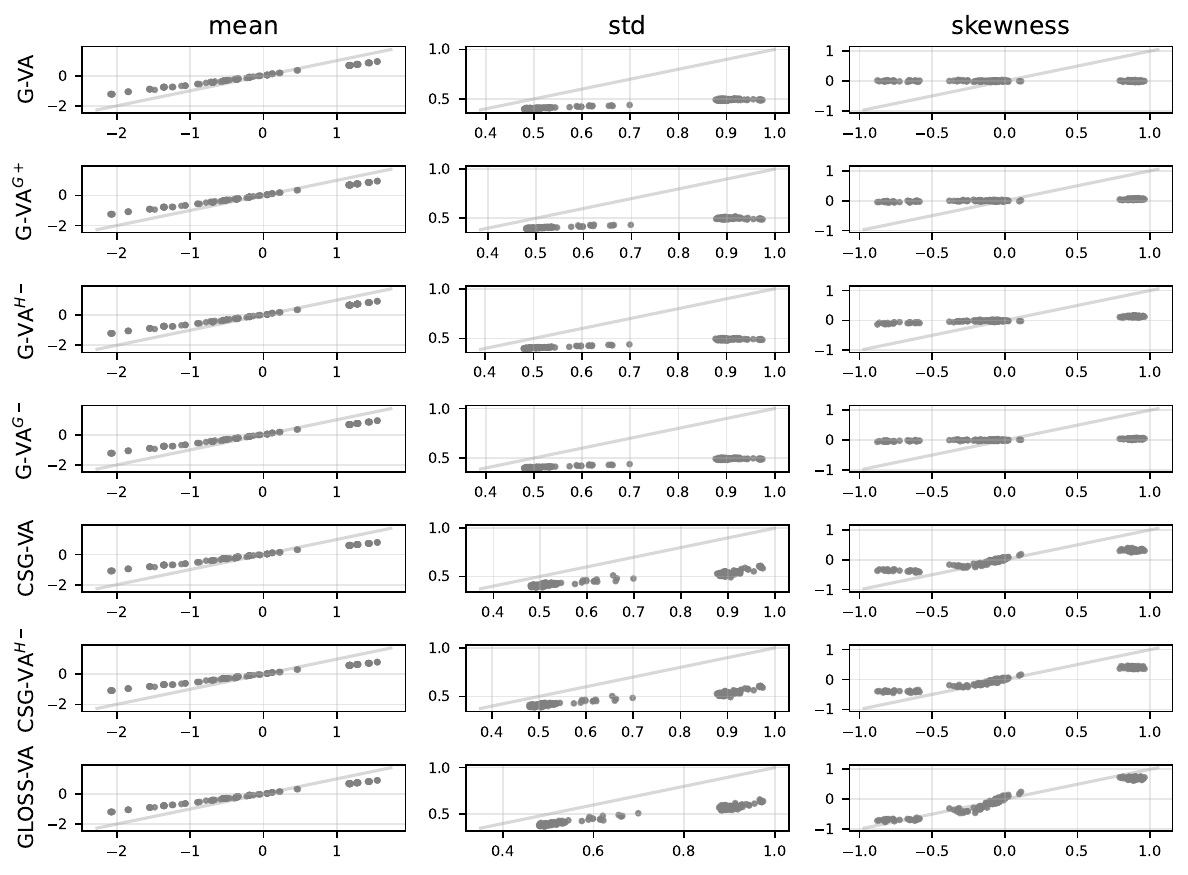}
    \caption{\small  MMNL model. Scatter plots for the mean (left), standard deviation (middle) and skewness (right) for the marginal posteriors of $b_{i,\texttt{td}}$ for each approximation method (columns) relative to estimates from MCMC.}
    \label{app:fig:parking_local2}
\end{figure}

\begin{figure}[tbh]
    \centering
    \includegraphics[width=0.95\linewidth,keepaspectratio]{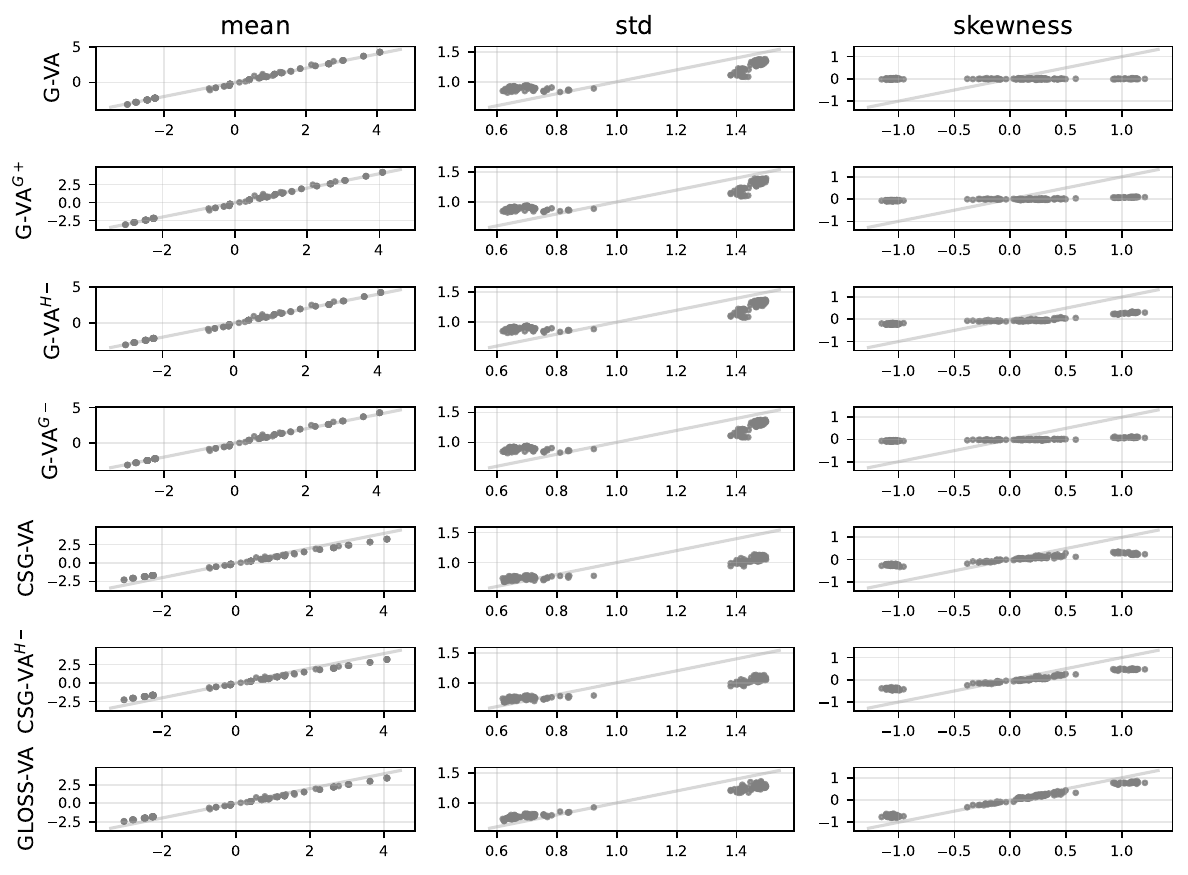}
    \caption{\small  MMNL model. Scatter plots for the mean (left), standard deviation (middle) and skewness (right) for the marginal posteriors of $b_{i,\texttt{fee}}$ for each approximation method (columns) relative to estimates from MCMC.}
    \label{app:fig:parking_local3}
\end{figure}

Figures~\ref{app:fig:parking_global} summarizes the marginal posterior approximations of the global variables for the MMNL model discussed in Section~3.3. Figures~\ref{app:fig:parking_local1}--\ref{app:fig:parking_local3} summarize the marginal posterior approximations of the latent variables. 


\subsection{Computational complexity}
\begin{figure}[htb!]
    \centering
    \includegraphics[width=0.95\linewidth,keepaspectratio]{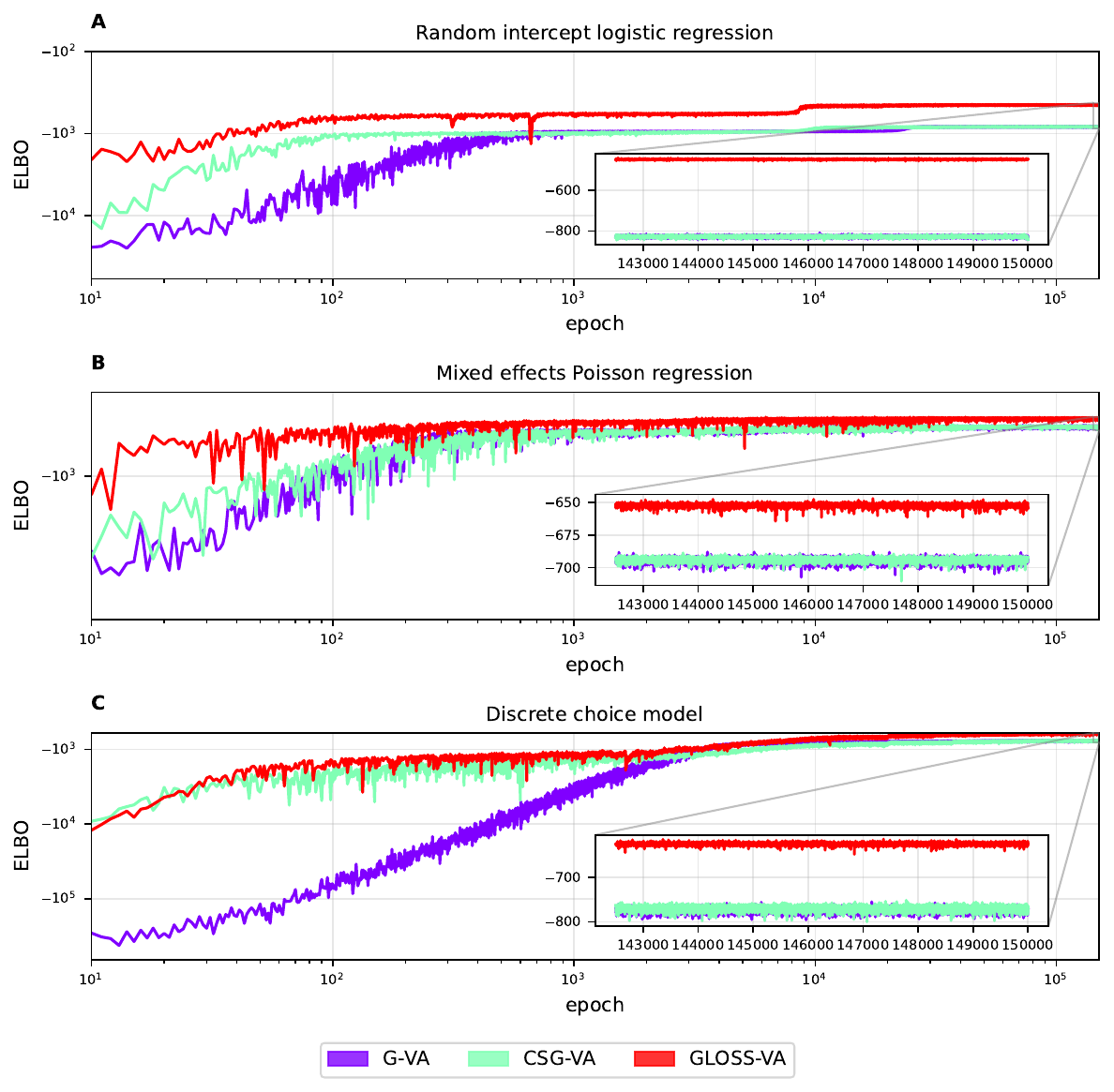}
    \caption{\small  Estimated ELBO versus iteration number for the logistic mixed model (\textbf{A}), Poisson mixed model (\textbf{B}) and discrete choice model (\textbf{C}) on log-scales.}
    \label{fig:elbos} 
\end{figure}

Table~\ref{app:tab:runtime} compares the runtime for the variational approximations and MCMC sampling across the three examples on a standard laptop. Note that applying the global or hierarchical skewness corrections post-hoc does not increase the runtime of the stochastic gradient ascent algorithm. MCMC samples are generated using RStan \citep{rstan} for the logistic mixed model, and the Poisson mixed model, while the MCMC sampler proposed by \citet{BanKruBieDazRas2020} is used for the discrete choice model.
Figure~\ref{fig:elbos} shows the estimated ELBO during training on a log-scale. 

\begin{table}[h!]
\centering
\begin{tabular}{c|cccc}
\hline
 {\bfseries Example} & {\bfseries G-VA} & {\bfseries CSG-VA} &  {\bfseries GLOSS-VA} & {\bfseries MCMC}\\
 \hline
Logistic mixed model &2.09&2.23&14.49&1.72\\
Poisson mixed model &2.11&2.22&18.48&4.47\\
Discrete choice model &6.68&7.03&65.26&2156.93\\
 \hline
\end{tabular}
\caption{\small Runtime in minutes.}
\label{app:tab:runtime}
\end{table}
\end{document}